\documentclass[11pt,a4paper]{article}
\usepackage[english]{babel}
\usepackage{jheppub}

\usepackage[usenames,dvipsnames,svgnames,table]{xcolor}

\usepackage{pdfpages}

\newcommand{\bfi}{\begin{figure}}
\newcommand{\efi}{\end{figure}}
\newcommand{\bc}{\begin{center}}
\newcommand{\ec}{\end{center}}
\newcommand{\beq}{\begin{equation}}
\newcommand{\eeq}{\end{equation}}
\newcommand{\beqn}{\begin{eqnarray}}
\newcommand{\eeqn}{\end{eqnarray}}

\title{Measurement of the neutrino velocity with the 
OPERA detector in the CNGS beam}

\author[a]{T.~Adam,}
\author[b]{N.~Agafonova,}
\author[c,1]{A.~Aleksandrov,}
\author[d]{O.~Altinok,}
\author[e]{P.~ Alvarez Sanchez,}
\author[f]{A.~Anokhina,}
\author[g]{S.~Aoki,}
\author[h]{A.~Ariga,}
\author[h]{T.~Ariga,}
\author[i]{D.~Autiero,}
% ...

\affiliation[a]{IPHC, Universit\'e de Strasbourg, CNRS/IN2P3, F-67037 Strasbourg, France}

\affiliation[b]{INR-Institute for Nuclear Research of the Russian Academy of Sciences, RUS-327312 Moscow, Russia}

\affiliation[c]{INFN Sezione di Napoli, I-80125 Napoli, Italy}

\affiliation[d]{METU-Middle East Technical University, TR-06532 Ankara, Turkey}

\affiliation[e]{European Organization for Nuclear Research (CERN), Geneva, Switzerland}

\affiliation[f]{(MSU SINP) Lomonosov Moscow State University Skobeltsyn Institute of Nuclear Physics, RUS-119992 Moscow, Russia}

\affiliation[g]{Kobe University, J-657-8501 Kobe, Japan}

\affiliation[h]{Albert Einstein Center for Fundamental Physics, Laboratory for High Energy Physics (LHEP), University of Bern, CH-3012 Bern, Switzerland}

\affiliation[i]{IPNL, Universit\'e Claude Bernard Lyon I, CNRS/IN2P3, F-69622 Villeurbanne, France}

% ...
\affiliation[1]{On leave of absence from LPI-Lebedev Physical Institute of the Russian Academy of Sciences, 119991 Moscow, Russia}

% ...

\clearpage

\abstract
{
The OPERA neutrino experiment at the underground Gran Sasso Laboratory has measured the velocity of neutrinos from the CERN CNGS beam over a baseline of about 730~km.
 The measurement is based on data taken by OPERA in the years 2009, 2010 and 2011. Dedicated upgrades of the CNGS timing system and of the OPERA detector, as well as a high precision geodesy campaign for the measurement of the neutrino baseline, allowed reaching comparable systematic and statistical accuracies. 
 
 An arrival time of CNGS muon neutrinos with respect to the one computed assuming the speed of light in vacuum of ($6.5 \pm 7.4~(stat.)^{+8.3}_{-8.0}  (sys.)$)~ns was measured corresponding to a relative difference of the muon neutrino velocity with respect to the speed of light  
 $(v-c)/c = (2.7 \pm 3.1~(stat.)~^{+3.4}_{-3.3}~(sys.)) \times 10^{-6}$. The above result, obtained by comparing the time distributions of neutrino interactions and of protons hitting the CNGS target in 10.5~$\mu$s long extractions, was confirmed by a test performed at the end of 2011 using  a short bunch beam allowing to measure the neutrino time of flight at the single interaction level.
  }


\clearpage

\begin{document}


\maketitle

\clearpage

\section{Introduction}
\label{sec1}

The OPERA neutrino experiment \cite{ref1} at the underground Gran Sasso Laboratory (LNGS) was designed to perform the first detection of neutrino oscillations in direct appearance mode in the $\nu_\mu \rightarrow \nu_\tau$ channel, the signature being the identification of the $\tau^-$ lepton created by its charged current (CC) interaction \cite{ref2}.

In addition to its main goal, the experiment is well suited to determine the neutrino velocity with high accuracy through the measurement of the time of flight and of the distance between the source of the CNGS neutrino beam at CERN (CERN Neutrino beam to Gran Sasso) \cite{ref3} and the OPERA detector at LNGS. For CNGS neutrino energies, $<~E_\nu~>~= 17$~GeV, the relative deviation from the speed of light $c$ of the neutrino velocity  due to its finite rest mass is expected to be smaller than $10^{-19}$, even assuming the mass of the heaviest neutrino \textit{eigenstate} to be as large as 2 eV \cite{ref4}. Hence, any larger deviation of the neutrino velocity, $v$, from $c$ would point to Lorentz invariance violation in the neutrino sector.

In the past, a high energy ($E_\nu > 30$~GeV) and short baseline experiment was able to test deviations down to $(v-c)/c < 4 \times 10^{-5}$ \cite{ref5}. With a baseline analogous to that of OPERA but at lower neutrino energies ($E_\nu$ peaking at ~3 GeV with a tail extending above 100 GeV), the MINOS experiment reported a measurement of $(v-c)/c = (5.1 \pm 2.9) \times 10^{-5}$ \cite{ref6}. At much lower energy, in the 10 MeV range, a stringent limit of $|v-c|/c < 2 \times 10^{-9}$ was set by the observation of (anti) neutrinos emitted by the SN1987A supernova \cite{ref7}.

In this paper we report on the determination of the neutrino velocity, defined as the ratio of the measured distance from CERN to OPERA to the time of flight of neutrinos traveling through the Earth's crust. We used the high-statistics data taken by OPERA in the years 2009, 2010 and 2011. Dedicated upgrades of the timing systems for the time tagging and synchronisation of the CNGS beam at CERN and of the OPERA detector at LNGS resulted in a reduction of the systematic uncertainties down to the level of the statistical error. The measurement also relies on a geodesy campaign that allowed measuring the 730~km CNGS baseline with a precision of 20~cm.

 Furthermore, in 2011 we conducted a measurement of the neutrino time of flight at the single interaction level  with a  short bunch beam, obtaining consistent results. Measurements obtained using two different OPERA subdetectors, namely the planes of plastic scintillator strips constituting the Target Tracker (TT) and Resistive Plate Chambers (RPC) are reported.

 The results presented in this paper were obtained by taking into account  the corrections for instrumental effects discovered after the originally reported neutrino velocity anomaly \cite{arxiv-v2}, see Section~\ref{sec6.1}.

\section{The OPERA detector and the CNGS neutrino beam}
\label{sec2}

The OPERA neutrino detector at LNGS is composed of two identical Super Modules, each consisting of an instrumented target section with a mass of about 625 tons followed by a magnetic muon spectrometre. Each section is a succession of walls filled with emulsion film/lead units interleaved with pairs of $6.7 \times 6.7$~m$^2$ planes of 256 horizontal and vertical scintillator strips composing the Target Tracker (TT). The TT allows the location of neutrino interactions in the target. This detector is also used to measure the arrival time of neutrinos. The scintillating strips are read out on both sides through wave length shifters (WLS) Kuraray Y11 fibres coupled to 64-channel Hamamatsu H7546 photomultipliers \cite{ref8}. Extensive information on the OPERA experiment is given in \cite{ref1} and in particular for the TT in \cite{ref9}.

\begin{figure}
  \centering
  \includegraphics[height=.50\textheight]{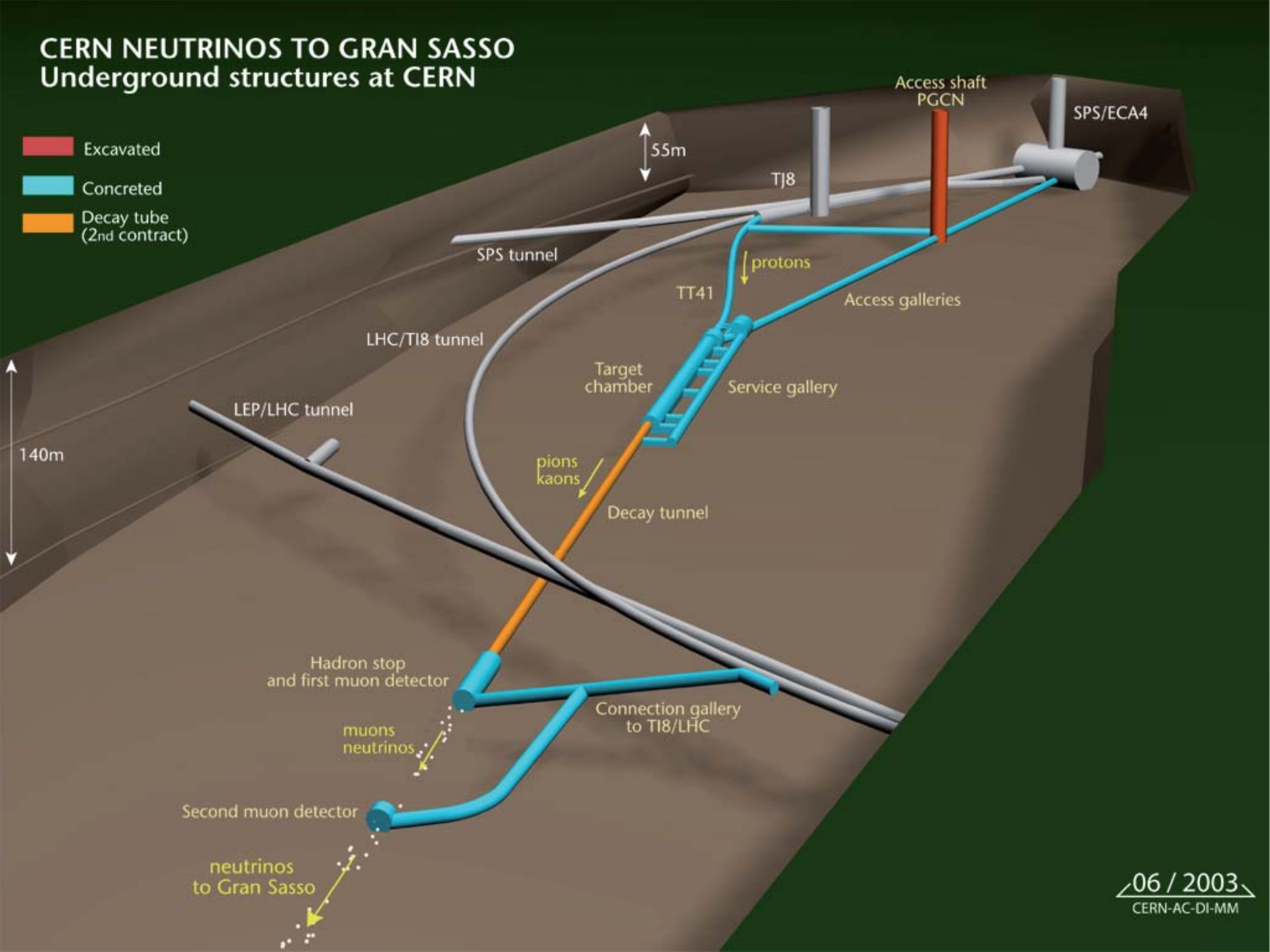}
  \caption{Artistic view of the SPS/CNGS layout.}
 \label{fig1}
\end{figure}

\begin{figure}
  \centering
  \includegraphics[height=.20\textheight]{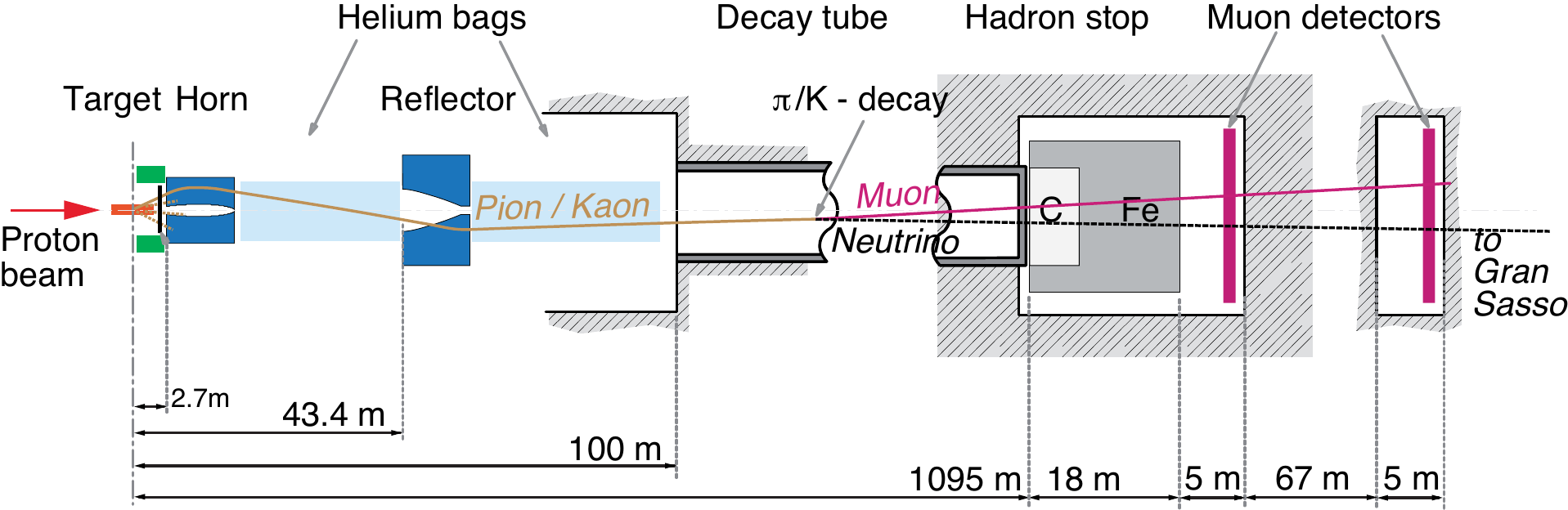}
  \caption{Layout of the CNGS beam line.}
 \label{fig2}
\end{figure}

The CNGS beam is produced by accelerating protons to 400 GeV/c with the CERN Super Proton Synchrotron (SPS). These protons are ejected with a kicker magnet towards a 2~m long graphite target in two extractions, each lasting 10.5 $\mu$s and separated by 50 ms. Each CNGS cycle in the SPS is 6 s long. Secondary charged mesons are focused by a magnetic horn and reflector, each followed by a helium bag to minimise the interaction probability of the mesons. These decay in flight, mainly into neutrinos and muons, in a 1000 m long evacuated tunnel. The SPS/CNGS layout is shown in Fig.~\ref{fig1}. The different components of the CNGS beam are shown in Fig.~\ref{fig2}.

The distance between the neutrino target and the OPERA detector is about 730 km. The CNGS beam is an almost pure $\nu_\mu$ beam with an average energy of 17 GeV, optimised for $\nu_\mu \rightarrow \nu_\tau$ appearance oscillation studies. In terms of interactions in the detector, the $\bar{\nu_\mu}$ contamination is 2.1\%, while $\nu_e$ and $\bar{\nu_e}$ contaminations are together smaller than 1\%.
The FWHM of the neutrino beam at the OPERA location is 2.8 km.

\begin{figure}
  \centering
  \includegraphics[height=.40\textheight]{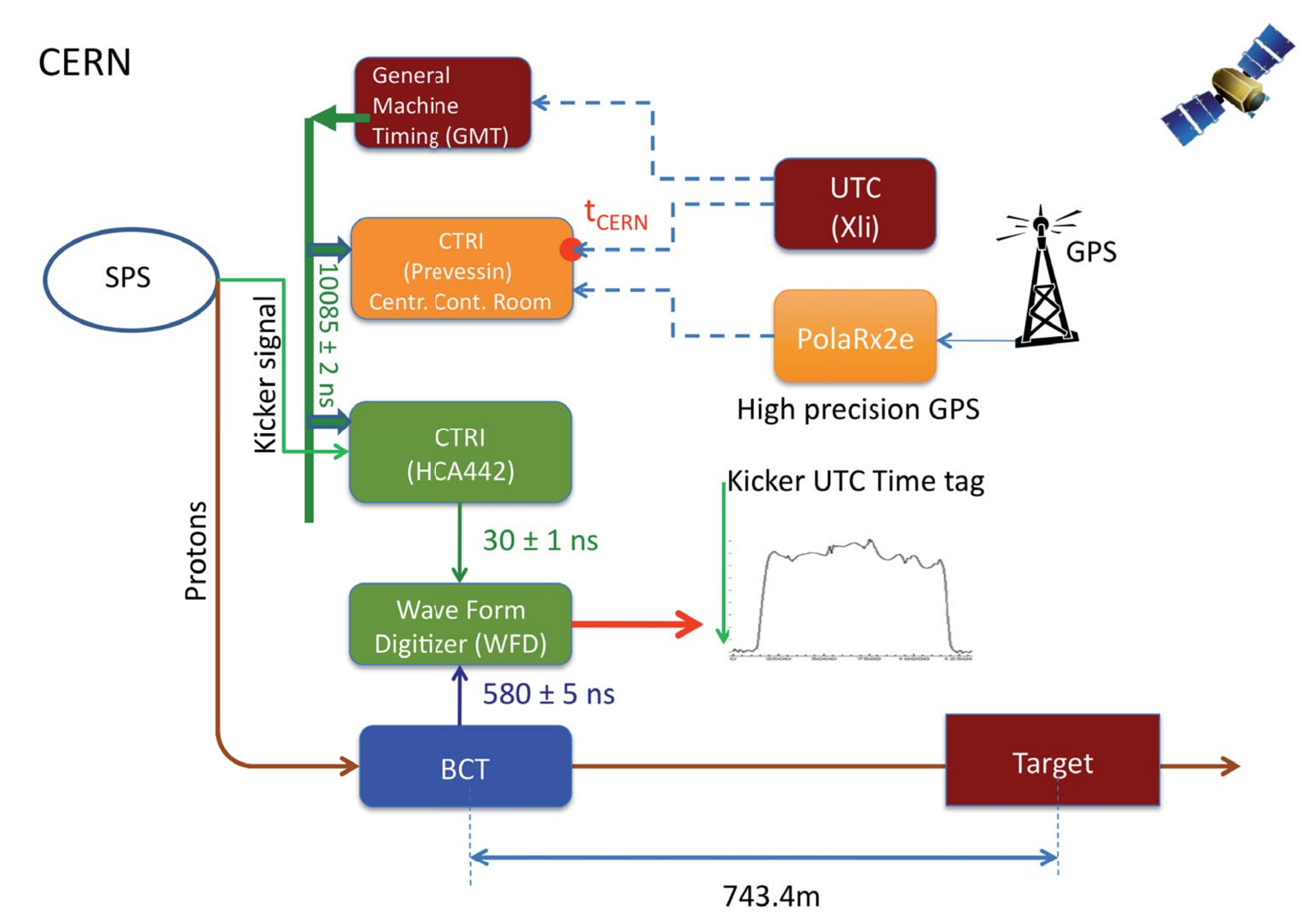}
  \caption{Schematic of the CERN SPS/CNGS timing system. Green boxes indicate detector time-response. Orange boxes refer to elements of the CNGS-OPERA synchronisation system. Details on the various elements are given in Section~\ref{sec6}.}
 \label{fig3}
   \vspace{1cm}
\end{figure}

The kicker magnet trigger-signal for the proton extraction from the SPS is UTC (Coordinated Universal Time) time-stamped with a Symmetricom Xli GPS receiver \cite{ref10}. The schematic of the SPS/CNGS timing system is shown in Fig.~\ref{fig3}. The determination of the delays shown in Fig.~\ref{fig3} is described in Section~\ref{sec6}.

The proton beam time-structure is accurately measured by a fast Beam Current Transformer (BCT) detector \cite{ref11} (BFCTI400344) located ($743.391 \pm 0.002$)~m upstream of the centre of the graphite target and read out by a 1~GS/s Wave Form Digitiser (WFD) Acqiris DP110 with a 250 MHz bandwidth \cite{ref12}. The BCT consists of toroidal transformers coaxial to the proton beam providing a signal proportional to the beam current transiting through it, with a 400 MHz
bandwidth. The linearity of the device is better than 1\% and it is operated far from the saturation limit. The start of the digitisation window of the WFD is triggered by the kicker magnet signal. The waveforms recorded for each extraction by the WFD are stamped with the UTC and stored in the CNGS database.

\begin{figure}
\begin{center}
\resizebox{!}{5.9cm}{\includegraphics{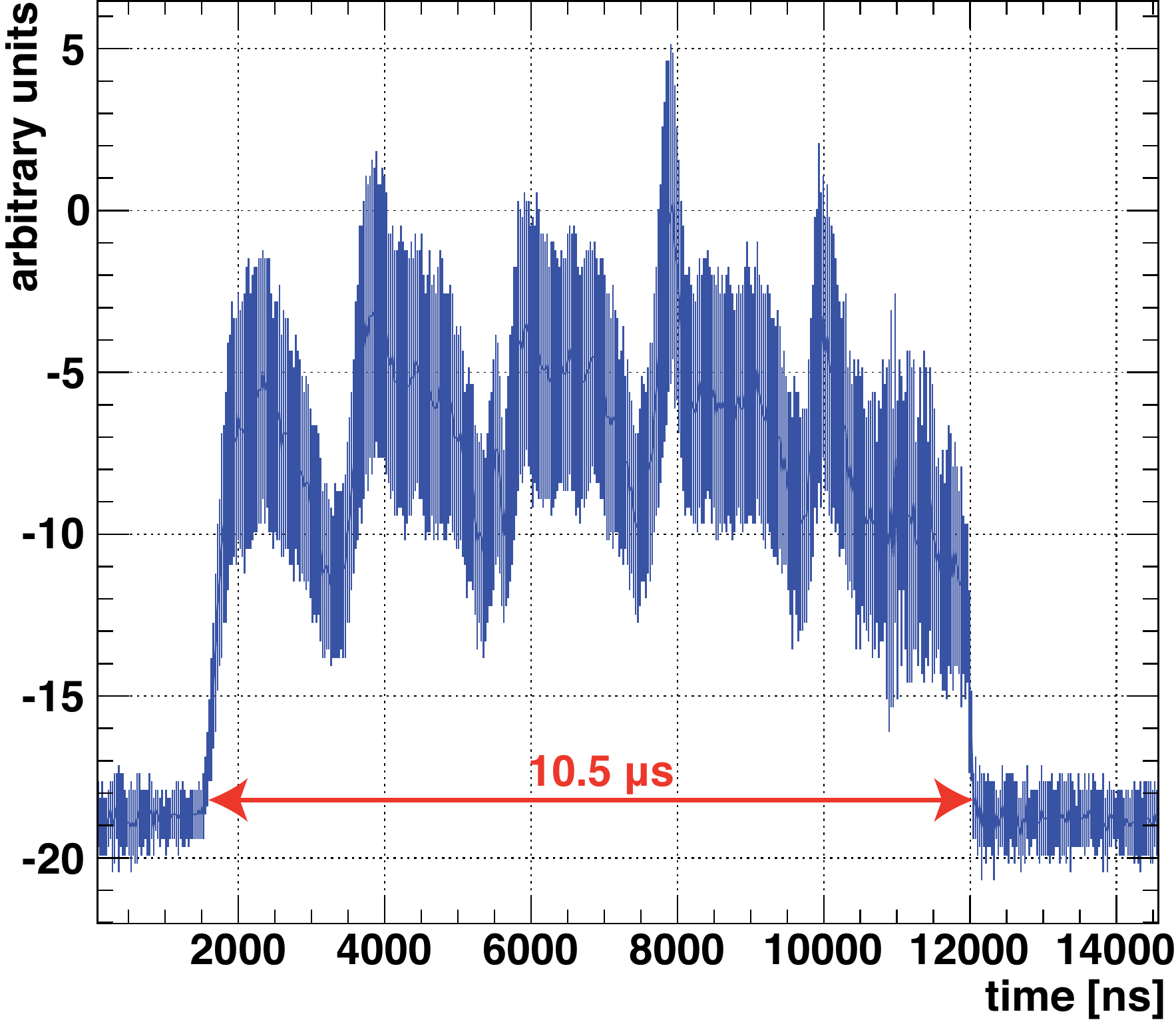}}
\hspace{0.5cm}
\resizebox{!}{5.9cm}{\includegraphics{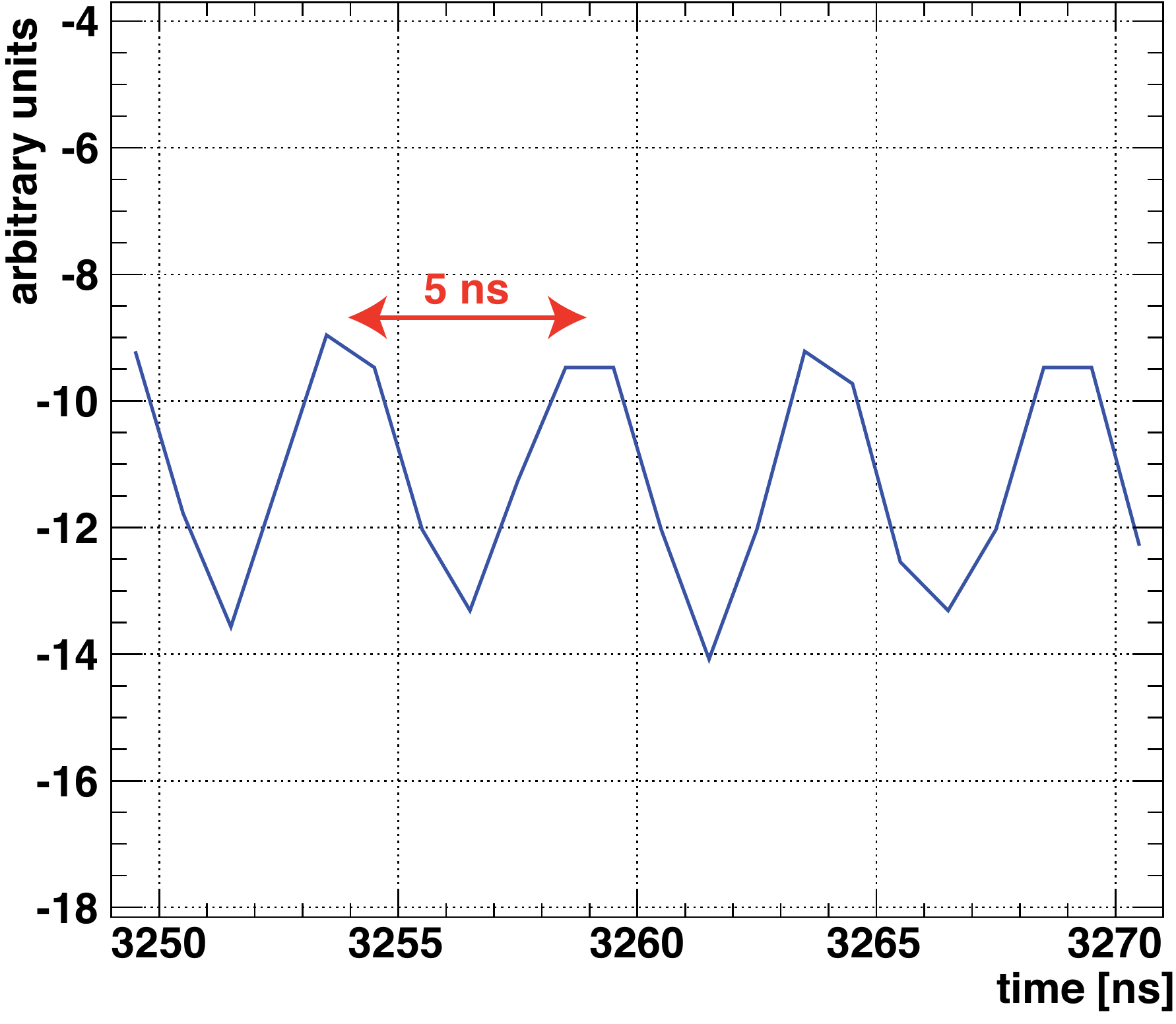}}
\end{center}
  \vspace{-0.2cm}
  \caption{Example of a specially selected proton extraction waveform measured with the BCT detector BFCTI400344 to show the five-peak structure reflecting the proton losses in the PS Continuous Transfer extraction mechanism. This structure is more pronounced than for the majority of the waveforms. A blow-up of the waveform (right plot) shows that the 200 MHz SPS radiofrequency is resolved.}

 \label{fig4}
\end{figure}

The intensity of the proton beam in the SPS features a five-step structure reflecting the five-turn (2.1 $\mu$s per turn) Continuous Transfer (CT) extraction mode from the CERN Proton Synchrotron (PS), as seen in the left part of Fig.~\ref{fig4}. The fine structure due to the 200 MHz SPS radiofrequency is superimposed, which is actually resolved by the BCT measurement (Fig.~\ref{fig4}, right).

\section{Principle of the neutrino time of flight measurement}
\label{sec3}

\begin{figure}
  \centering
  \includegraphics[height=.50\textheight]{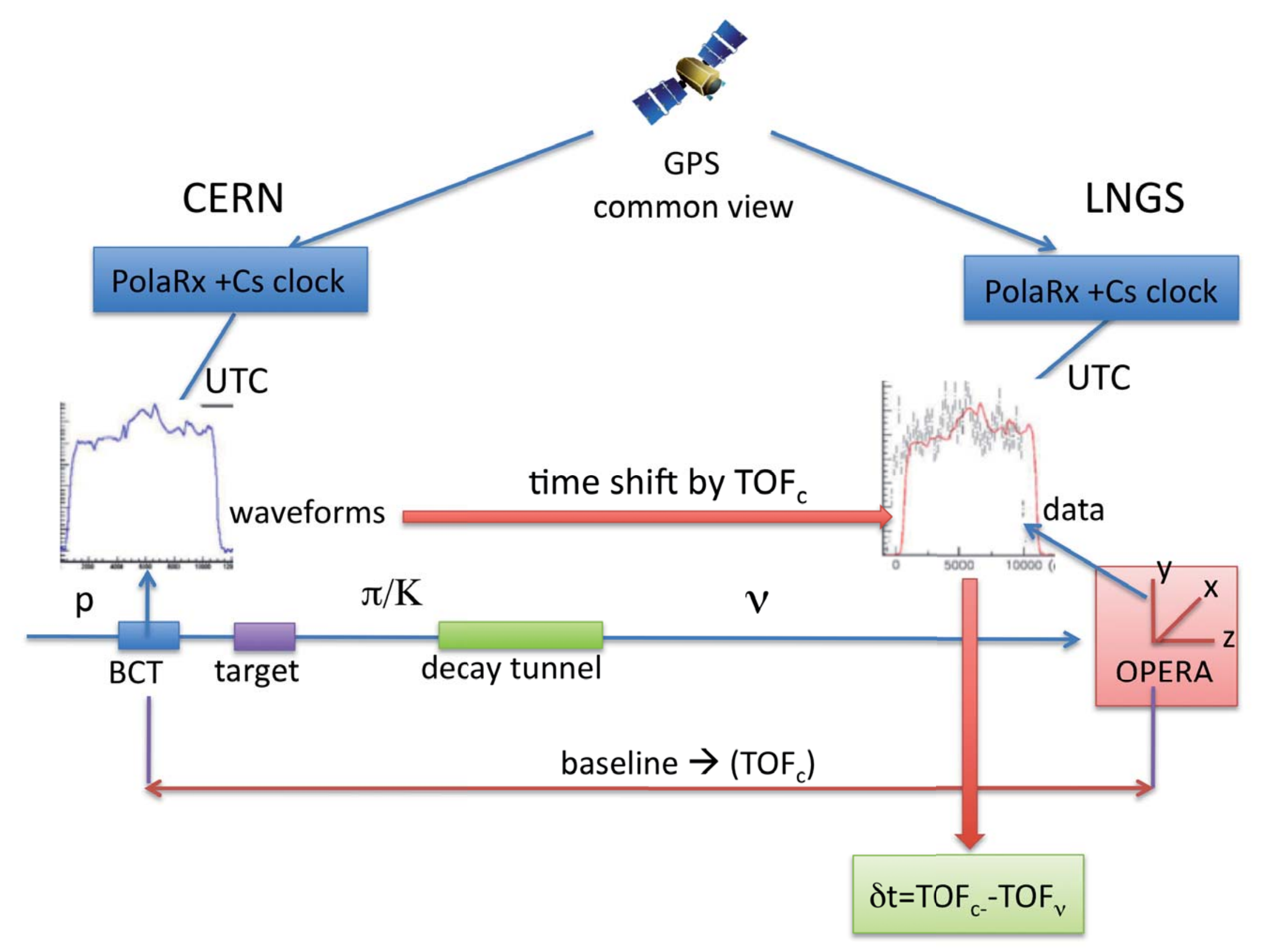}
  \caption{Schematic of the time of flight measurement.}
 \label{fig5}
\end{figure}

A schematic description of the principle of the time of flight measurement is shown in Fig.~\ref{fig5}. The time of flight of CNGS neutrinos ($TOF_\nu$) cannot be precisely measured at the single interaction level since any proton in the 10.5 $\mu$s extraction time may produce the neutrino detected by OPERA. However, by measuring the time distributions of protons for each extraction for which neutrino interactions are observed in the detector, and summing them together, after proper normalisation one obtains the probability density function (PDF) of the time of emission of the neutrinos within the duration of extraction. Each proton waveform is UTC time-stamped as well as the events detected by OPERA. The two time-stamps are related by $TOF_c$, the expected time of flight assuming the speed of light \cite{ref13}. It is worth stressing that this measurement does not rely on the difference between a start and a stop signal but on the comparison of two event time distributions.

The PDF distribution can then be compared with the time distribution of the interactions detected in OPERA, in order to measure $TOF_\nu$. The deviation $\delta t = TOF_c - TOF_\nu$ is obtained by a maximum likelihood analysis of the time tags of the OPERA events with respect to the PDF, as a function of ${\delta t}$. The individual measurement of the waveforms reflecting the time structure of the extraction reduces systematic effects related to time variations of the beam compared to the case where the beam time structure is measured on average, e.g. by a near neutrino detector without using proton waveforms.

The total statistics used for the analysis reported in this paper is 15223 events detected in OPERA, corresponding to about $10^{20}$ protons on target collected during the 2009, 2010 and most of the 2011 CNGS runs. This allowed estimating $\delta t$ with a small statistical uncertainty, presently comparable to the total systematic uncertainty.

\begin{figure}
  \centering
  \includegraphics[height=.50\textheight]{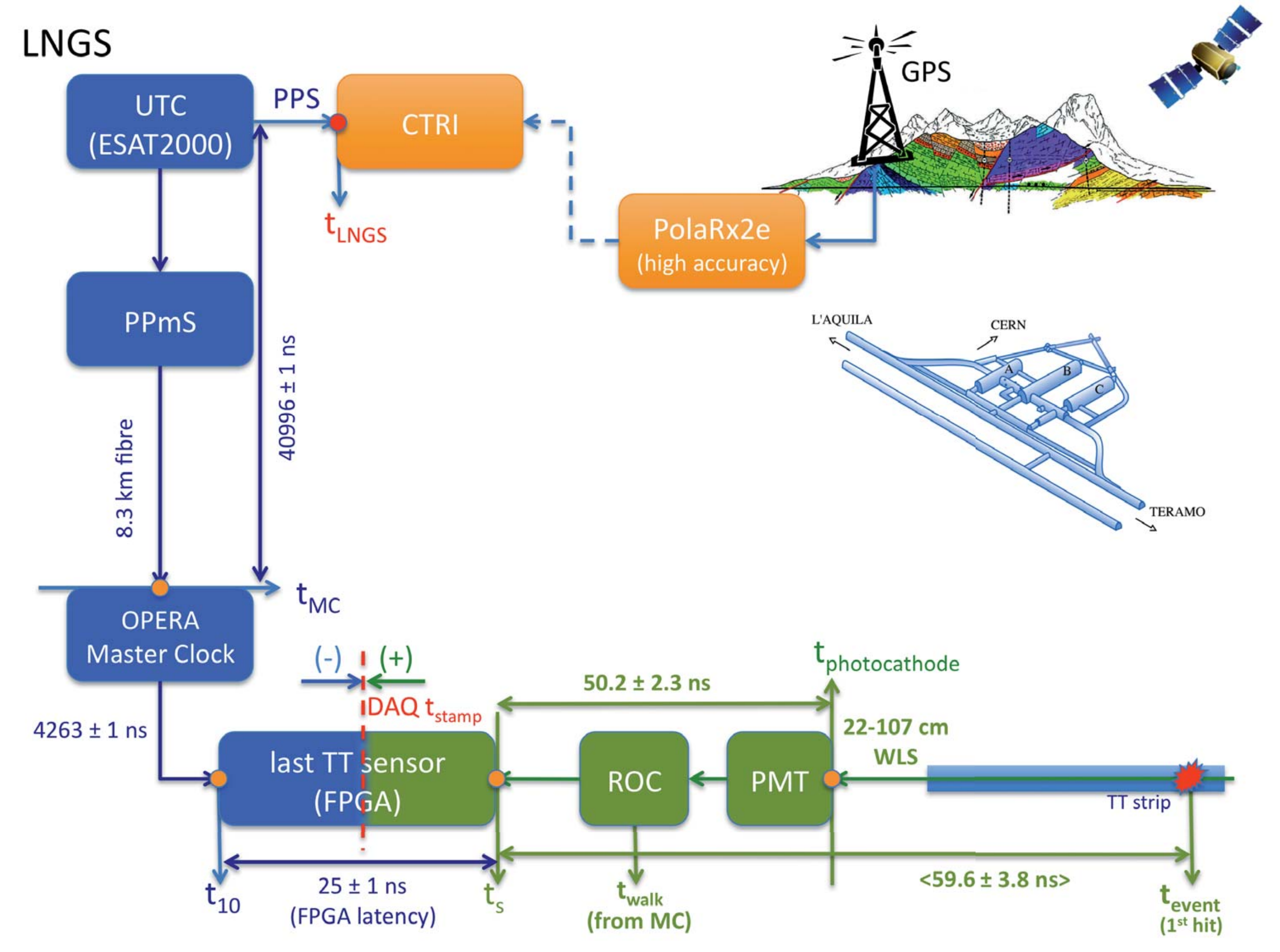}
  \caption{Schematic of the OPERA timing system at LNGS. Blue delays include elements of the time-stamp distribution; increasing delays decrease the value of $\delta t$. Green delays indicate detector time-response; increasing delays increase the value of $\delta t$. Orange boxes refer to elements of the CNGS-OPERA synchronisation system.}
 \label{fig6}
\end{figure}

The reference point used for the baseline measurement at CERN is the position of the BCT which is 743.4 m upstream of the target. Assuming that 400 GeV protons travel at the speed of light over this distance induces a negligible correction of 0.007 ns. The point where the parent meson produces a neutrino in the decay tunnel is unknown. However, this introduces a negligible inaccuracy in the neutrino time of flight measurement, because the produced mesons are ultra-relativistic. This affects $TOF_\nu$ by a correction of the order of $d/2c\gamma^2$, with $d$ being the meson decay length, on average 450~m from the target, and $\gamma$ its Lorentz factor, on average as large as 190. By a full FLUKA-based simulation of the CNGS beam \cite{ref14} it was shown that the time difference computed assuming a particle moving at the speed of light from the target down to LNGS, with respect to the value derived by taking into account the speed of the relativistic parent meson down to its decay point is less than $0.02$ ns. Similar arguments apply to muons produced in muon neutrino CC interactions occurring in the rock in front of the OPERA detector and seen in the apparatus (external events). With a full GEANT simulation of external events it was shown that ignoring the position of the interaction point in the rock introduces a bias smaller than 2 ns with respect to those events occurring in the target (internal events), provided that external interactions are selected by requiring identified muons in OPERA. More details on the muon identification procedure are given in \cite{ref15}.

A key feature of the neutrino velocity measurement is the accuracy of the relative time tagging at CERN and at the OPERA detector. The standard GPS receivers formerly installed at CERN and LNGS would feature an insufficient 100~ns accuracy for the $TOF_\nu$ measurement. Thus, in 2008, two identical systems, composed of a GPS receiver for time-transfer applications Septentrio PolaRx2e \cite{ref16} operating in ``common-view'' mode \cite{ref17} (a method in which only signals emitted by GPS satellites seen simultaneously by the receivers at both ends of the baseline are considered), and a Cs atomic clock Symmetricom Cs4000 \cite{ref18} were installed at CERN and LNGS (see Figs.~\ref{fig3}, \ref{fig5} and \ref{fig6}).

The Cs4000 oscillator provides the reference frequency to the PolaRx2e receiver, which is able to time-tag its ``One Pulse Per Second'' output (1PPS) with respect to the individual GPS satellite observations. The latter are processed offline by using the CGGTTS format \cite{ref19}. The two systems feature a technology routinely used for high-accuracy time-transfer applications by national time and frequency metrology laboratories around the world, in order to compare atomic clocks \cite{ref20}.
These international time comparisons are the basis of the UTC as defined by the Bureau International des Poids et Mesures (BIPM). The two systems were calibrated in 2008 by the  Swiss Federal Metrology Institute METAS (Bundesamt f\"{u}r Metrologie) \cite{ref21} and established a permanent time link at the 2~ns level between two reference points ($t_{CERN}$ and $t_{LNGS}$) of the timing chains at CERN and OPERA. This time link was independently verified in 2011 by the  German Federal Metrology Institute PTB (Physikalisch-Technische Bundesanstalt) \cite{ref22} by taking data at CERN and LNGS with a portable time-transfer device commonly employed for relative time link calibrations \cite{ref23}. The difference between the time base of the CERN and OPERA PolaRx2e receivers was measured to be ($2.3 \pm 0.9$)~ns \cite{ref22}. This correction was taken into account in the application of the time link.

All the other elements of the timing distribution chains of CERN and OPERA were determined using different techniques, further described in the following, aiming to reach a comparable level of accuracy.

\section{Measurement of the neutrino baseline}
\label{sec4}

The other fundamental ingredient for the neutrino velocity measurement is the knowledge of the distance between the point where the proton time-structure is measured at CERN and the origin of the reference frame for the OPERA underground detector at LNGS. The relative positions of the elements of the CNGS beam line are known with millimetre accuracy. When these coordinates are transformed into the global geodesy reference frame ETRF2000~\footnote{The International Terrestrial Reference System (ITRS) is a set of procedures defined by the International Union of Geodesy and Geophysics; it allows to  determine the coordinates of fixed points on the surface of the Earth with a cm precision as well as their drift velocities. The European Terrestrial Reference Frame (ETRF) is a particular realisation of the ITRS for measuring relative coordinates within Europe where the Eurasian plate is taken as static.} \cite{ref24} by relating them to external GPS benchmarks  at CERN, they are known within 2 cm accuracy. This frame has a scale error at the level of $10^{-9}$ \cite{ref25}.

The analysis of the GPS benchmark positions was first done by extrapolating measurements taken at different periods via geodynamical models \cite{ref26}, and then by comparing simultaneous measurements taken in the same reference frame. The two methods yielded the same result within 2~cm \cite{ref25}. The travel path of protons from the BCT to the focal point of the CNGS target is also known with millimetre accuracy.

The distance between the target focal point and the OPERA reference frame was precisely measured in 2010 following a dedicated geodesy campaign. The coordinates of the origin of the OPERA reference frame were measured by establishing GPS benchmarks at the two sides of the 10~km long Gran Sasso highway tunnel and by transporting their positions with a terrestrial traverse down to the OPERA detector. A common analysis in the ETRF2000 reference frame of the 3D coordinates of the OPERA origin and of the target focal point allowed the determination of this distance to be ($730534.61 \pm 0.20$)~m \cite{ref25}. The 20~cm uncertainty is dominated by the 8.3~km underground link between the outdoor GPS benchmarks and the benchmark at the OPERA detector \cite{ref25}.

\begin{figure}
  \centering
  \includegraphics[height=.50\textheight]{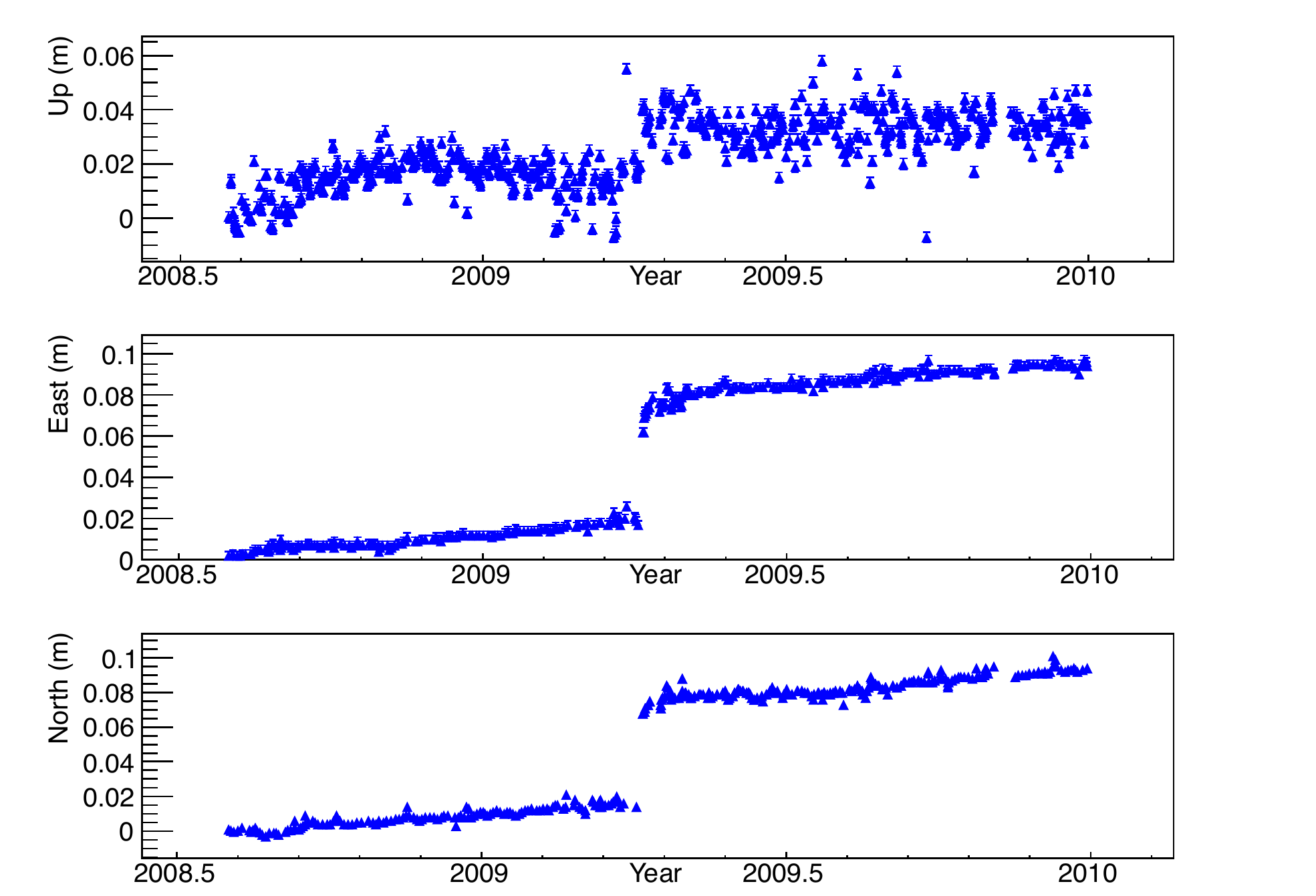}
  \caption{Monitoring of the PolaRx2e GPS antenna position at LNGS showing the slow earth crust drift and the fault displacement due to the 2009 earthquake in the L'Aquila region. Units for the horizontal (vertical) axis are years (metres).}
 \label{fig7}
\end{figure}

The accurate time-transfer GPS receiver PolaRx2e allows to continuously monitor tiny movements of the Earth's crust, such as continental drift that shows up as a smooth variation of less than 1~cm/year, and the detection of slightly larger effects due to earthquakes. The April 2009 earthquake in the region of LNGS, in particular, produced a sudden displacement of about 7~cm, as seen in Fig.~\ref{fig7}. All mentioned effects are within the accuracy of the baseline determination.

Tidal effects occurring during the geodesy measurements were corrected for by expressing the results in a conventional tide-free frame \cite{ref27}. Therefore, measurements taken at different times can be directly compared. As far as the neutrino baseline is concerned, periodic tidal movements are below the 1~cm level and are averaged over the long data-taking period \cite{ref25}.

The baseline considered for the measurement of the neutrino  time of flight 
is then the sum of ($730534.61 \pm 0.20$)~m between the CNGS target focal point and the origin of the OPERA detector reference frame, and ($743.391 \pm 0.002$)~m between the BCT and the focal point, \textit{i.e.} ($731278.0 \pm 0.2$)~m.

\section{Data selection}
\label{sec5}

The OPERA data acquisition system (DAQ) time-tags the  TT detector hits with 10~ns quantization with respect to the UTC \cite{ref28}. The time of a neutrino interaction is defined as that of the earliest hit in the TT. CNGS events are preselected by requiring that they fall within a window of $20~\mu$s with respect to the SPS kicker magnet trigger-signal, delayed by the neutrino time of flight assuming the speed of light and corrected for the various delays of the timing systems at CERN and at OPERA. The relative fraction of cosmic ray events accidentally falling in this window is $10^{-4}$ and it is therefore negligible~\cite{ref1,ref29}.

Since $TOF_c$ is computed with respect to the origin of the OPERA reference frame, located beneath the most upstream spectrometre magnet, the time of the earliest hit for each event is corrected for its distance along the beam line from this point, assuming time propagation at the speed of light. The UTC time of each event is also individually corrected for the instantaneous value of the time link correlating the CERN and OPERA timing systems, as obtained from the two PolaRx2e receivers. These corrections reflect the instability of the standard GPS systems at CERN and LNGS, whose time bases may vary by several tens of nanoseconds with respect to each other on a few hours scale.

The total statistics used for this analysis (15223 events) includes 7235 internal (charged and neutral current interactions) and 7988 external (charged current) events. Internal events, preselected by the electronic detectors with the same procedure used for neutrino oscillation studies \cite{ref30} constitute a subsample of the entire OPERA statistics  accumulated during the considered run time (about 70\%), for which both time transfer systems at CERN and LNGS were operational, as well as the database-logging of the proton waveforms. As mentioned before, external events, in addition, are requested to have a muon identified in the detector.

The final statistics of 15223 neutrino interactions does not include about 5\% of the preselected events, characterized by an earliest hit isolated in time and in position inside the detector with respect to the bulk of the event hits, which were discarded. Such isolated hits may be due to noise not included in the simulations and therefore constitute a potential source of bias towards early arrival times. For the retained events there is a good agreement between data and simulations as far as the timing of the earliest hit is concerned. This is discussed in the next section.

\section{Neutrino event timing}
\label{sec6}

The schematic of the SPS/CNGS timing system is shown in Fig.~\ref{fig3}. A general-purpose timing receiver ``Control Timing Receiver'' (CTRI) at CERN \cite{ref31} logs every second the difference in time between the 1PPS outputs of the Xli and of the more precise PolaRx2e GPS receivers, with 0.1 ns resolution. The Xli 1PPS output represents the reference point of the time link to OPERA. This point is also the source of the ``General Machine Timing'' chain (GMT) serving the CERN accelerator complex \cite{ref32}.

The GPS devices are located in the CERN Pr\'evessin Central Control Room (CCR). The time information is transmitted via the GMT to a remote CTRI device in Hall HCA442 used to UTC time-stamp the kicker magnet signal. This CTRI also produces a delayed replica of the kicker magnet signal, which is sent to the adjacent WFD module. The UTC time-stamp marks the start of the digitisation window of the BCT signal. The latter signal is brought via a coaxial cable to the WFD at a distance of 100 m. Three delays characterise the CERN timing chain:

\begin{enumerate}

\item {The propagation delay through the GMT of the time base of the CTRI module logging the PolaRx2e 1PPS output to the CTRI module used to time-tag the kicker pulse $\Delta t_{UTC } = (10085 \pm 2$)~ns;}

\item {The delay to produce the replica of the kicker magnet signal from the CTRI to start the digitisation of the WFD $\Delta t_{trigger} = (30 \pm 1$)~ns;}

\item {The delay from the time the protons cross the BCT to the time a signal arrives to the WFD $\Delta t_{BCT} = (580 \pm 5$)~ns.}

\end{enumerate}

The kicker signal is used as a pre-trigger and as an arbitrary time origin. The measurement of the $TOF_\nu$ is based instead on the BCT waveforms, which are tagged with respect to the UTC.

The measurement of $\Delta t_{UTC}$ was performed by means of a portable Cs4000 oscillator. Its 1PPS output, stable to better than 1ns over a few hour scale, was input to the CTRI used to log the Xli 1PPs signal at the CERN CCR. The same signal was then input to the CTRI used to time-stamp the kicker signal at the HCA442 location. The two measurements allowed the determination of the delay between the time bases of the two CTRI, and to relate the kicker time-stamp to the Xli output. The measurements were repeated three times during the last two years and yielded the same results within 2 ns. This delay was also determined by performing a two-way timing measurement with optical fibres. The Cs clock and the two-way measurements also agree within 2~ns.

The two-way measurement is a technique routinely used in this analysis for the determination of delays. Measuring the delay $t_A$ in propagating a signal to a far device consists in sending the same signal via an optical fibre $B$ to the far device location in parallel to its direct path $A$. At this site the time difference $t_A-t_B$ between the signals following the two paths is measured. A second measurement is performed by taking the signal arriving at the far location via its direct path $A$ and sending it back to the origin with the optical fibre $B$. At the origin the time difference between the production and receiving time of the signal corresponds to $t_A+t_B$. In this procedure the optoelectronic chain used for the fibre transmission of the two measurements is kept identical by simply swapping the receiver and the transmitter between the two locations. The two combined measurements allow determining $t_A$ \cite{ref33}. For the $\Delta t_{UTC }$ the two-way setup was left in operation since July 2011 to assess the time stability of the results. The measured $\Delta t_{UTC}$ showed excursions not exceeding 0.4~ns related to temperature variations of the 2~km long fibres and of the associated electronics.

Measurements by two-way fibre and transportable Cs clock were systematically compared for the determination of the various delays of the CERN and OPERA timing chains and agreed within 1~ns. The two techniques are based on an inclusive measurement of the delay between pairs of reference points. This does not introduce any bias that could be related to  the calibration of individual hardware elements of the chain.

$\Delta t_{trigger}$ was estimated by an accurate oscilloscope measurement. The determination of $\Delta t_{BCT}$ was first performed by measuring the 1PPS output of the Cs4000 oscillator with a digital oscilloscope and comparing it to a CTRI signal at the point where the BCT signal arrives at the WFD. This was compared to a similar measurement where the Cs4000 1PPS signal was injected into the calibration input of the BCT. The time difference of the 1PPS signals in the two configurations led to the measurement of $\Delta t_{BCT }= (581 {\pm} 10)$~ns.

\begin{figure}
  \centering
  \includegraphics[height=.45\textheight]{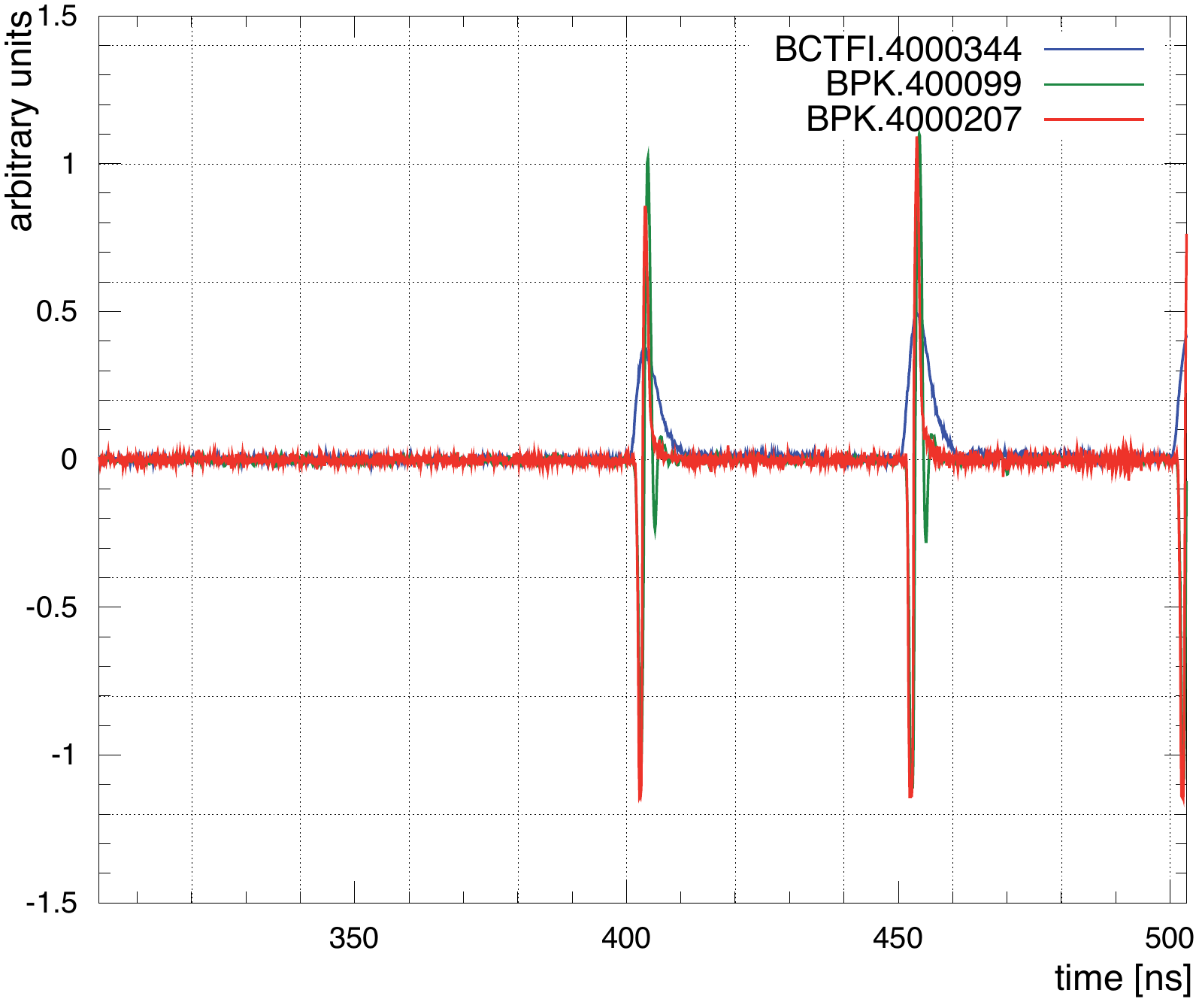}
  \caption{Comparison between the signals of the BCT and of the fast pick-up detectors after
compensating for $\Delta t_{BCT}$.}
 \label{fig8}
\end{figure}

Since the above determination through the calibration input of the BCT might not be representative of the internal delay of the BCT with respect to the transit time of the protons, and also because the error on this measurement was by far the largest contribution to the overall systematic uncertainty, a more sophisticated method was then applied. The proton transit time was tagged upstream of the BCT by two fast beam pick-ups BPK400099 and BPK400207 with a time response of 1~ns \cite{ref34}. From the relative positions of the three detectors (the pick-ups and the BCT) along the beam line and the signals from the two pick-ups one determines the time the protons cross the BCT and the time delay at the level of the WFD. In order to achieve an accurate determination of the delay between the BCT and the BPK signals, a measurement was performed in the particularly clean experimental condition of the SPS proton injection to the Large Hadron Collider (LHC) machine of 12 bunches with a width of about 1 ns and with 50 ns spacing, passing through the BCT and the two pick-up detectors. This measurement was performed simultaneously for the 12 bunches and yielded $\Delta t_{BCT} = (580 \pm 5~(sys.))$~ns. The systematic error also accounts for uncertainties on the modelling of the time response of the BCT, including cables and electronics, which results in a broadening of the digitised signal with respect to the proton current pulse. This is illustrated in Fig.~\ref{fig8} for proton bunches of 1 ns.


The schematic of the OPERA timing system at LNGS is shown in Fig.~\ref{fig6}. The official UTC time source at LNGS is provided by a GPS system ESAT 2000 \cite{ref35,ref36} operating at the surface laboratory. The 1PPS output of the ESAT is logged with a CTRI module every second with respect to the 1PPS of the PolaRx2e, in order to establish a high-accuracy time link with CERN. Every millisecond a pulse synchronously derived from the 1PPS of the ESAT (PPmS) is transmitted to the underground laboratory via an 8.3 km long optical fibre. The delay of this transmission with respect to the ESAT 1PPS output down to the OPERA Master Clock output was measured with a two-way fibre procedure in July 2006 and amounts to ($40996 \pm 1$)~ns. Additional measurements with a transportable Cs clock were also performed in June 2007 yielding the same result.
As we will see in Section~\ref{sec6.1}, a series of additional measurements on the OPERA timing system conducted during the 2011 CNGS winter shut down led to an effective fibre delay during the whole data taking period 2008-2011 of 41069 ns. 
Based on the dispersion of the measured values during this period (after August 2008 until December 2011, see Fig.~\ref{fig6.1}), a systematic uncertainty of 3.7 ns on the  fibre delay was estimated.


The OPERA Master Clock is disciplined by a  Vectron OC-050 oscillator with an Allan deviation of $2\times 10^{-12}$ at an observation time of 1~s. This oscillator keeps the local time during the 0.6  s DAQ cycle. The OPERA Master Clok is synchronised at every DAQ cycle start  with the PPmS signal coming from the external GPS. This signal is tagged with respect to the uncorrelated internal frequency  producing a $\pm~25$~ns time jitter (this jitter is only relevant for the low statistics run with the bunched beam, see Section~\ref{sec9}). The frequency of the Vectron oscillator was measured during the 2011 CNGS shut down and found slightly larger ($0.124$ ppm) than specified (see Section~\ref{sec6.1}).

The time base of the OPERA Master Clock is transmitted to the frontend cards of the TT with the FPGA (see Fig.~\ref{fig6}).  This delay ($\Delta t_{clock}$) was also measured with two techniques, namely by the two-way fibre method, and by transporting the Cs4000 clock to the two points. Both measurements provided the same result of ($4263 \pm 1$)~ns. The frontend card time-stamp is performed in a Field Programmable Gate Array (FPGA) by incrementing a coarse counter every 0.6~s and a fine counter with a frequency of 100~MHz. At the occurrence of a trigger the content of the two counters provides a measure of the arrival time. The fine counter is reset every 0.6 s by the arrival of the Master Clock signal that also increments the coarse counter. The internal delay of the FPGA processing the Master Clock signal to reset the fine counter was determined by a parallel measurement of trigger and clock signals with the DAQ and a digital oscilloscope. This measured delay (FPGA latency) is ($24.5 \pm 1.0$)~ns. This takes into account the 10~ns quantization effect due to the clock period.

The delays in producing the Target Tracker signal including the scintillator response, the propagation of the signals in the WLS fibres, the transit time of the photomultiplier \cite{ref8}, and the time response of the OPERA analogue frontend readout chip (ROC) \cite{ref37} were inclusively calibrated by exciting the scintillator strips at known positions by a UV picosecond laser \cite{ref38}. The arrival time distribution of the photons  at  the photocathode and the time walk due to the discriminator threshold in the analogue frontend chip as a function of the signal pulse height were accurately determined in laboratory measurements and included in the detector simulation. The total time elapsed from the moment photons reach the photocathode, a trigger is issued by the ROC analogue frontend chip, and the trigger signal arrives at the FPGA where it is time-stamped, was  ($50.2 \pm 2.3$)~ns.

Since the time response to neutrino interactions depends on the position of the hits in the detector and on their pulse height, the average TT delay was evaluated from the difference between the exact interaction time and the time-stamp of the earliest hit for a sample of fully simulated neutrino interactions. Starting from the position at which photons are generated in each strip, the simulation takes into account all the effects determined in laboratory measurements including the arrival time distribution of the photons for a given production position, the time-walk of the ROC chip, and the measured delays from the photocathode to the FPGA. This TT delay has an average value of 59.6~ns with a RMS of 7.3~ns, reflecting the transverse event distribution inside the detector. The 59.6~ns represents the overall delay of the TT response down to the FPGA and it includes the quoted delay of 50.2~ns.    The simulation procedure adds a 3~ns term to the systematic error.

\begin{figure}
\begin{center}
\resizebox{!}{5.1cm}{\includegraphics{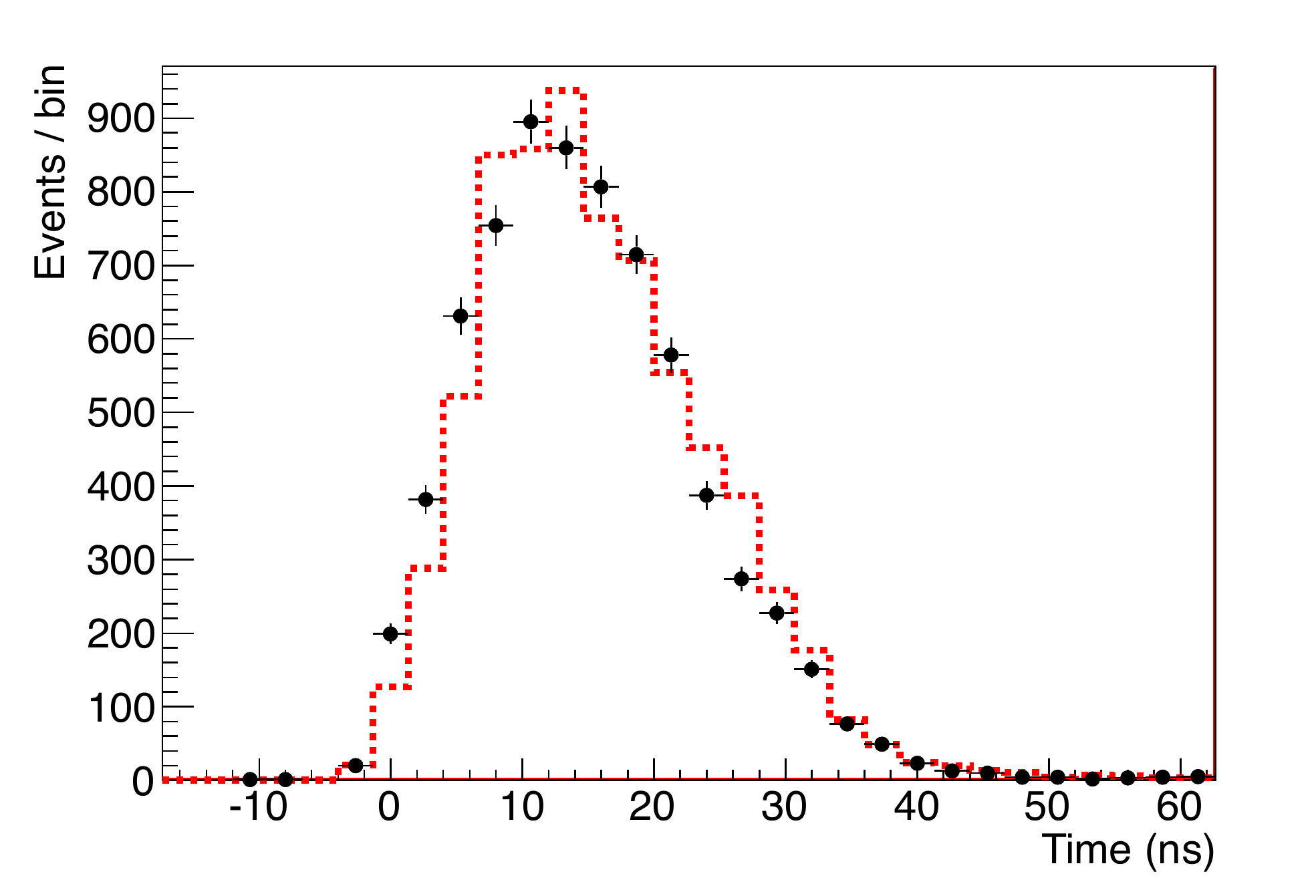}}
\resizebox{!}{5.1cm}{\includegraphics{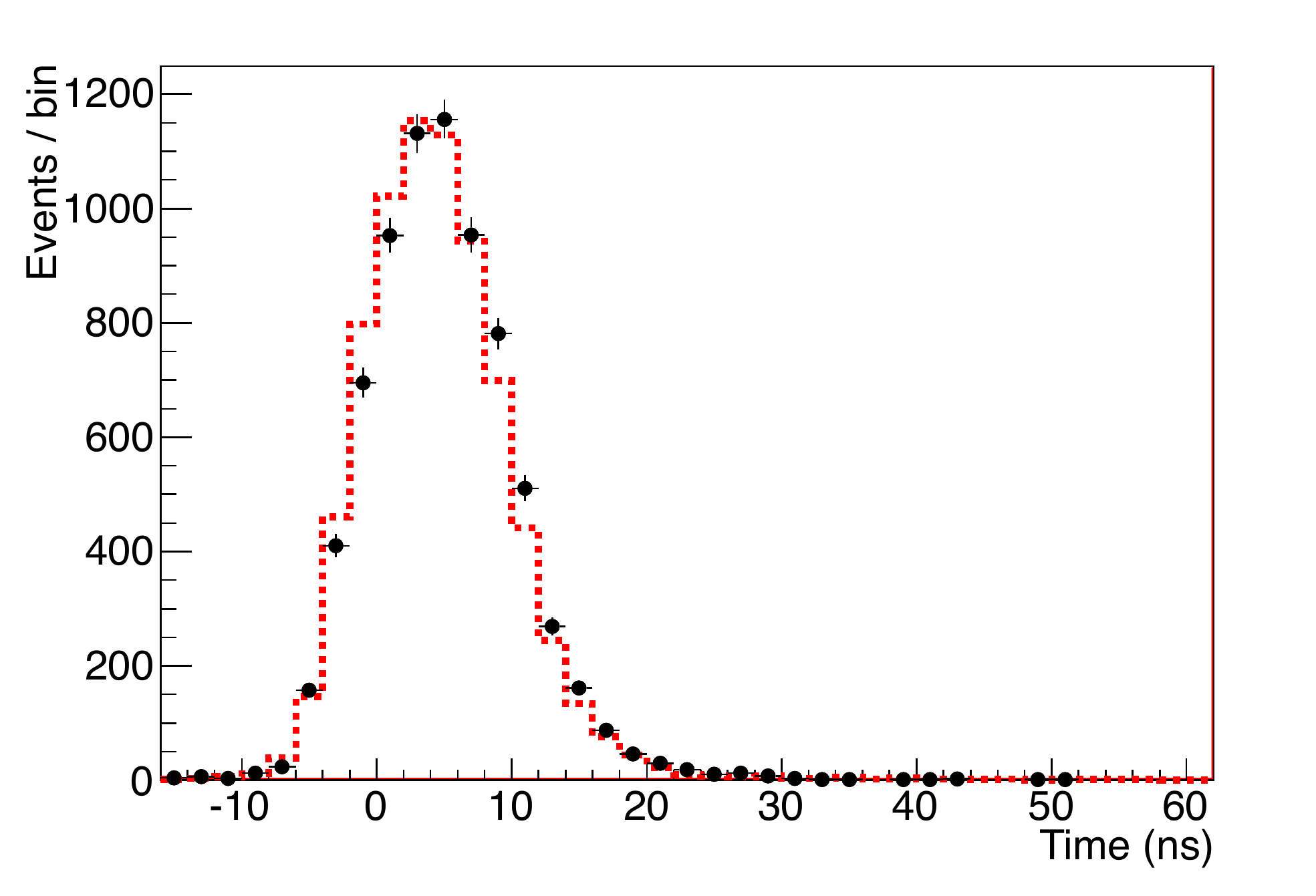}}
\end{center}
\vspace{-0.4cm}
  \caption{Distribution of the time difference between the earliest TT hit and: a) the average time of the event, b) the average time of the muon track. Dots with error bars indicate data and the dotted line simulated events. Plot a) includes only internal events while plot b) only external events. The distributions are corrected for the longitudinal position of the hits.}
 \label{fig9}
\end{figure}

Several checks were performed by comparing data and simulated events as far as the earliest TT hit timing is concerned. Data and simulations agree within the above-mentioned systematic uncertainty of 3~ns for both the time differences between the earliest and all the following hits, and for the difference between the earliest hit and the average timing of muon tracks. This is shown in Fig.~\ref{fig9} where the distribution of the time difference between the earliest TT hit and the average time of the event, and the average time of the muon track are shown for internal and external events, respectively. The distributions are corrected for the longitudinal position of the hits. Consequently, after correction, the truly earliest hit used to time the event may appear to be preceded by more downstream hits, hence the negative value occasionally taken by the time difference.

Corrections were also applied to take into account the Sagnac effect caused by the rotation of the Earth around its axis. This yields an increase of $TOF_c$ by 2.2~ns, with a negligible error. The Earth's revolution around the Sun and the movement of the solar system in the Milky Way induce a negligible effect, as well as the influence of the gravitational fields of Moon, Sun and Milky Way, and the Earth's frame-dragging \cite{ref39}. The relative effect of the Earth's gravitational field on the Schwarzschild geodesic amounts to $10^{-8}$ and it is therefore totally negligible. The gravitational red-shift due to the different CERN and LNGS altitudes produces an even smaller relative effect of $10^{-13}$ on the clocks in between two common-view synchronisations~\cite{ref39}.

More details on the neutrino timing and on the geodesy measurement procedures can be found in \cite{ref40}.

\subsection{Measurements performed during the 2011 CNGS winter shut down }
\label{sec6.1}
At the end of the 2011 CNGS run the OPERA timing at LNGS was further checked in order to test its stability over time ~\cite{ref400}. 



Measurements were made, starting at the beginning of December 2011, of the time delay in the 8.3 km optical fibre  between the ESAT GPS 1PPS output and  the OPERA Master Clock output using the standard 2-way technique. 
A value $73.2$~ns larger than the one determined in 2006 and 2007 and a larger jitter of the Opera Master Clock  latching of the GPS signal were measured. Further investigations, that lasted until mid February 2012, revealed that the difference originated from an optical cable not properly connected  thus reducing  the amount of light received by the optical/electrical converter of the Master Clock. When proper connections were restored, 
during a technical intervention, the values of the delay  and of the jitter were found to agree with what  was measured in 2006 and 2007, as listed below:\\

\begin{tabular} {lrc}
2006 	&  $t_A = ( 40995.5 \pm 0.3$) ns & RMS $=~3.2$~ns\\ 
2011: before fibre reconnection &	 $t_A = ( 41068.6 \pm 0.5)$ ns &  RMS $=~6.0$~ns \\
2011: after fibre reconnection &	 $t_A = ( 40994.1 \pm 0.3)$ ns	 & RMS $=~3.2$~ns\\
\end{tabular}

\begin{figure}
  \centering
  \includegraphics[height=0.13\textheight]{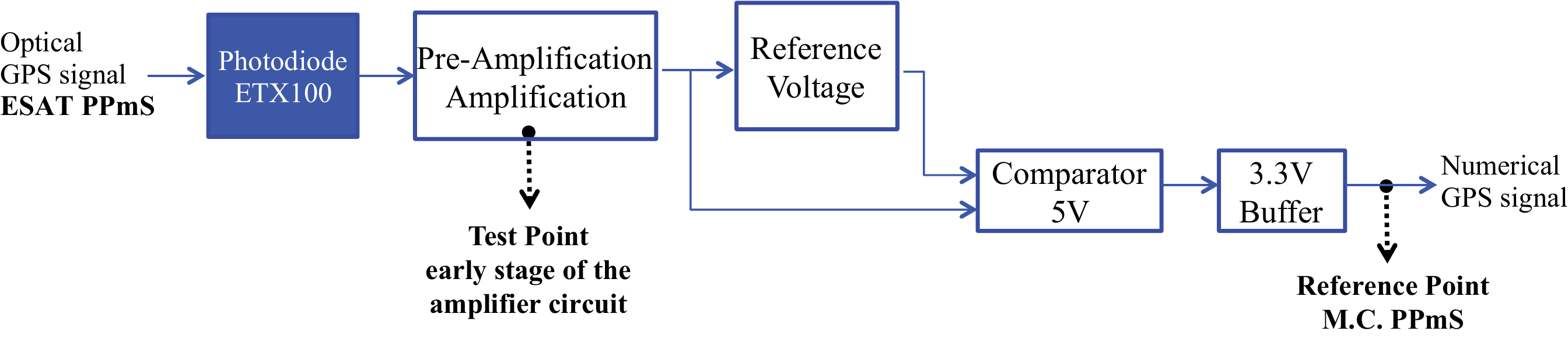}
  \caption{A simplified scheme of the OPERA Master Clock opto-electronic circuit used to convert the ESAT PPmS optical signal into an electric signal.}
 \label{fig6.00}
\end{figure}

\begin{figure}
  \centering
  \includegraphics[height=.37\textheight]{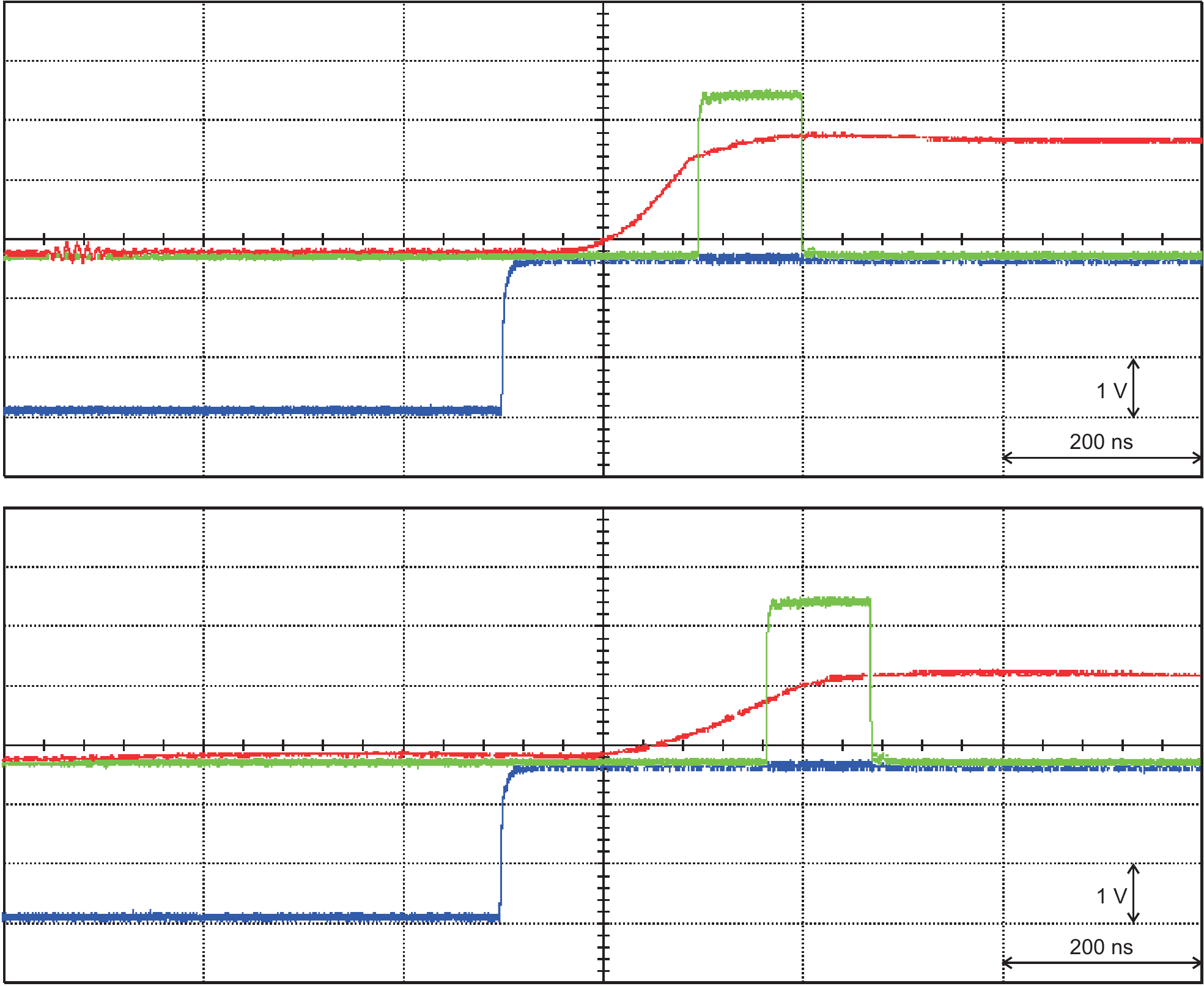}
  \caption{The ESAT 1PPS time reference propagated underground through an 8.3 km optical path (blue), the Master Clock PPmS (green) and the Master Clock PPmS taken at an early stage of the amplifier circuit (red). Top: signals taken with the connectors correctly plugged; bottom: signals taken with connectors wrongly screwed in positions which provide an extra delay of $\sim$$74$ ns.}
 \label{fig6.0}
\end{figure}

\paragraph{}
Additional tests showed that the fibre delay could vary according to the amplitude of the light signal at the Master Clock  input (Fig.~\ref{fig6.00}). By acting on the optical fibre connections both rise time and plateau of the amplifier output signal varied depending on the input light intensity (Fig.~\ref{fig6.0}).  The effect was related to the  slow  electronics and the time-walk of the comparator.

The reset signal sent to the TT front end is delayed by the same amount as the Master Clock PPmS, leading to an underestimation of the neutrino TOF.

 During the CNGS winter shut~down the delay  $\Delta t_{clock}$  from the OPERA Master Clock to the last TT sensor  (FPGA) of the first TT wall   (taken as time reference)  was  re-measured. The replica of the 1PPmS ESAT signal  generated by the Master Clock (Fig.~\ref{fig6.00})  taken as time reference   was transmitted through an extra (calibrated) path and compared  to the  signal at point t$_{10}$ in Fig.~\ref{fig6}. A delay of  4262 ns was obtained confirming the previous measurement.

An additional test used the  same extra path and 1PPmS signal  to inject   a signal equivalent to 100 p.e. at the PMT connector level by means of a 2.3~pF capacitor.  The analysis of the time-stamps of such generated events, recorded  by the DAQ  as  it is done for  neutrino data taking, showed that the time delay between  two consecutive events (separated by 1 ms) was 0.124 ns larger than expected. 
As a consequence the event time stamp was overestimated by a quantity which depends on the event position inside the DAQ cycle, up to a maximum of 74 ns.
This effect was due to a frequency offset between the OPERA Master Clock and ESAT GPS2 oscillators, as proven   by comparing them to  a Cs Frequency Standard (10 MHz). The 10 MHz Vectron OC-050 Oscillator of the OPERA Master Clock  frequency  was slightly larger (0.124 ppm) than the specifications.



An additional independent information about the value of the LNGS fibre delay  during the 2009-2011 neutrino data taking came from an analysis tool developed to study  cosmic muon events in delayed coincidence in the OPERA and LVD detectors  as a signature of high $p_T$  events in cosmic rays \cite{ref402}. The analysis had revealed no such events but confirmed the existence of a significant flux of almost horizontal cosmic muons from the so called ``Teramo Valley'', a  region in the massif orography with large zenith angles ($\theta>80 \deg$) and modest rock thickness ($\sim$$2200$ m)~\cite{macro}. The idea was to use these data to determine when the above mentioned discrepancy in the fibre delay at LNGS occurred or started to develop. Since single horizontal muons traverse first the OPERA detector and then, at a distance of $\sim$160 m, the LVD apparatus,  any change in the timing chain of one of the two experiments would be reflected in a change of the cosmic muon time of flight  over  the OPERA-LVD distance. 
The main results of the analysis from mid 2007 until March 2012, are  reported in~\cite{ref402}.  

\begin{figure}
\centering
 \includegraphics[height=.4\textheight]{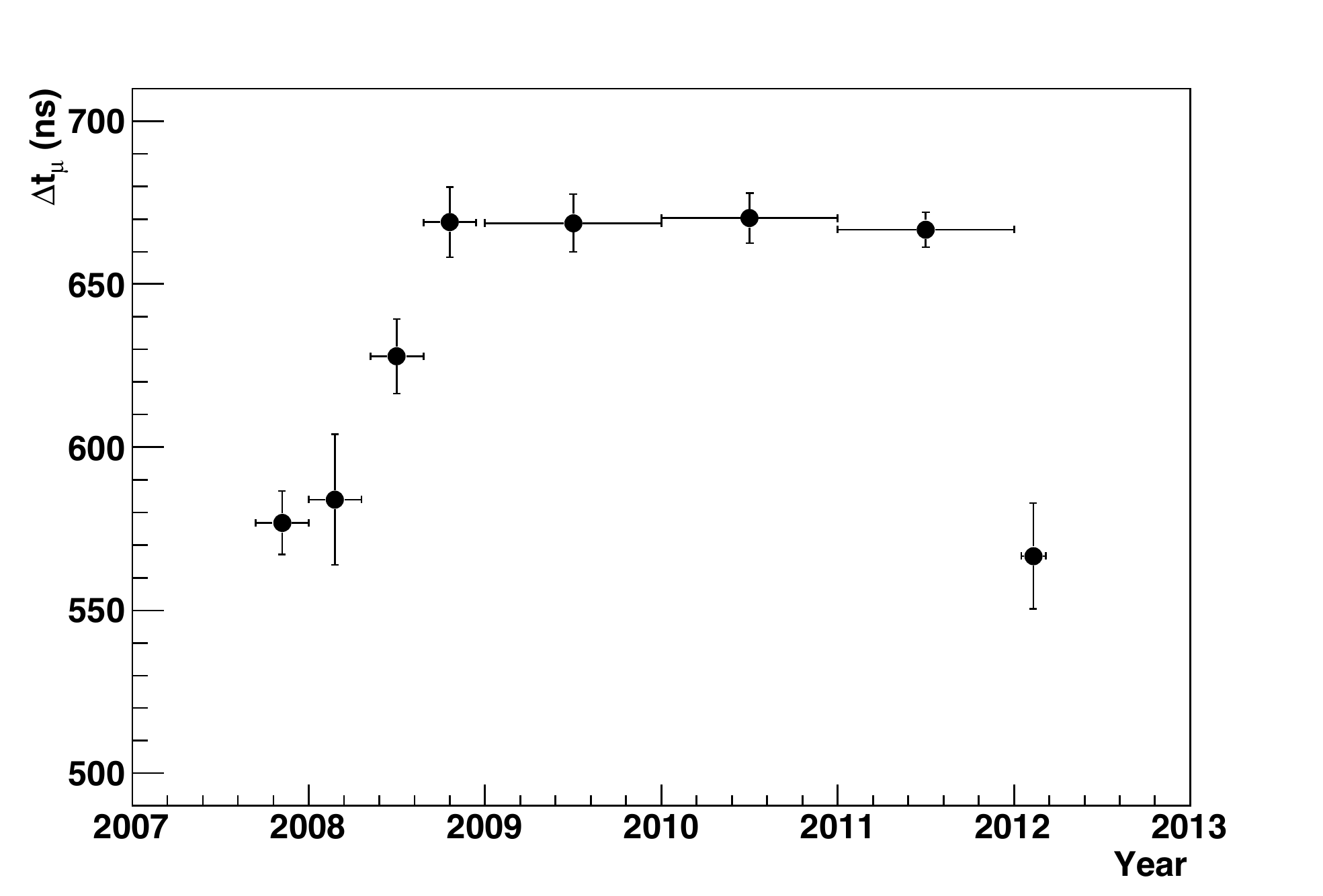}
  \caption{ Distribution of the $\Delta t_\mu = t_{LVD} - t_{OPERA}$. For  each year all events
are grouped into one single point, but for 2008 they are subdivided into three periods:  before May,  May-August, and after August.}
 \label{fig6.1}
\end{figure}

 In Fig.~\ref{fig6.1}   (taken from \cite{ref402})  the local time difference, $\Delta t_\mu$, between the muon recorded by OPERA and the same muon recorded by LVD is plotted versus time. 
 The figure indicates two abrupt changes in $\Delta t_\mu$, one around  August 2008 and the other in December 2011.  In between  $\Delta t_\mu$ stays constant thus corresponding to a  stable configuration of the detectors' timing systems. 
The extracted value of  (73.2 $\pm$~9)~nsec \cite{ref402} is compatible with the fibre recalibration described in Section~\ref{sec6.1}. The  observed  $\Delta t_\mu$ decrease in December 2011 is related to proper reconnection of  the optical fibre  to the OPERA Master Clock  during a technical intervention.  The
increase of  $\Delta t_\mu$ in August 2008 originated from an anomalous set-up of the OPERA timing
 system which remained in a  stable configuration over the whole data taking period
 considered in the neutrino velocity analysis. The data dispersion of Fig.~\ref{fig6.1} compared
 to a horizontal line in the flat region between 2008 and 2011 corresponds to 3.7 ns. This dispersion has been considered as the systematic error on the 8.3 km fibre delay.
 

 As indicated in \cite{ref402}, by comparing the OPERA  time stamps as a function of the time within the OPERA DAQ cycle,  a constant time drift of (114~$\pm$~14)~ns/s  was observed,   compatible with the previously reported value of 124 ns/s.  Grouping data year by year, this drift  remains constant within the errors. The 14 ns uncertainty on the  drift was used to derive the systematic error on the $\Delta t_{clock}$ delay between the Master Clock and the TT sensors, i.e. (4262~$\pm$ 2)~ns.

\section{Data analysis}
\label{sec7}
The  parameters used to compute $TOF_\nu$ and yielding the final value of $\delta t = TOF_c -TOF_\nu$ are summarised in Table~\ref{tab1}.
 In order to ease the interpretation of the corrections a sign is attributed to each calibration value:  delays  increasing (decreasing) the value of $\delta t$  have a positive (negative) sign.

\begin{table}
  \centering
\caption{Summary of the time delay values used in the analysis.}
\begin{tabular}{lr}
\hline
 & \bf {Time delay (ns)} \\
\hline
Baseline   & 2439280.9 \\
Earth rotation (Sagnac effect)                       & 2.2 \\
&\\
\bf{UTC corrections at CERN:}               &  \\
CTRI signal propagation through GMT chain $\Delta t_{UTC}$              & 10085.0 \\

Kicker magnet signal to WFD $\Delta t_{trigger}$           & 30 \\

BCT signal to WFD $\Delta t_{BCT}$          & -580 \\
 & \\
\bf{UCT corrections at Gran Sasso:}               &  \\
LNGS 8.3 km fibre to OPERA Master Clock& -41068.6\\
TT response  to FPGA         & 59.6 \\
 FPGA latency          & -24.5 \\
Master Clock to FPGA $\Delta t_{clock}$           & -4262.9 \\
 & \\
\bf{GPS Corrections:}               &  \\
 Time-link           & -2.3 \\
 & \\
\hline
\end{tabular}
\label{tab1}
\end{table}

For each neutrino interaction measured in the OPERA detector the analysis procedure used the corresponding proton extraction waveform. These waveforms were individually normalised to unity and summed up in order to build a PDF w(t). The noise present at the level of individual waveforms (see the baseline of the pulses shown in Fig.~\ref{fig4}) is averaged out by summing them up. The 200 MHz radiofrequency structure is still present in the final PDF, together with a coherent noise affecting its central part. This noise is due to an electromagnetic disturbance of the electronics in Hall HCA442, occurring with a constant delay with respect to the kicker magnet pulse. The same was observed independently by a different WFD operating in parallel and reading out another BCT detector. The  same noise was observed during empty spills where SPS protons are not sent to the CNGS line but to the beam dump. Since this noise is not related to the proton beam, it was filtered out by a low-pass filter applied to the final PDF. Checks were performed throughout the whole analysis chain to ensure that the filtering procedure did not affect the final results.

The WFD is triggered by the kicker magnet pulse, but the time of the proton pulses with respect to the kicker trigger is different for the two extractions. In fact, for the second extraction the kicker magnet pulse is anticipated with respect to the proton bunches, profiting of the fact that the SPS ring is half-empty. The kicker trigger is just related to the pulsing of the kicker magnet. The exact timing of the proton pulses stays within this large window of the pulse.

A separate maximum likelihood procedure was then carried out for the two proton extractions. The likelihood to be maximised for each extraction is a function of the single variable $\delta t$ to be added to the time tags $t_j$ of the OPERA events. These are expressed in the time reference of the proton waveform digitiser ($w_k$) assuming neutrinos traveling at the speed of light, such that their distribution best coincides with the corresponding PDF:

\begin{equation}
L_k ( \delta t_k ) = \prod_j W_k ( t_j + \delta t_k ) \ \ \  k  =  1,2 \rm{\ extractions}
\end{equation}

\begin{figure}
  \centering
  \includegraphics[height=.45\textheight]{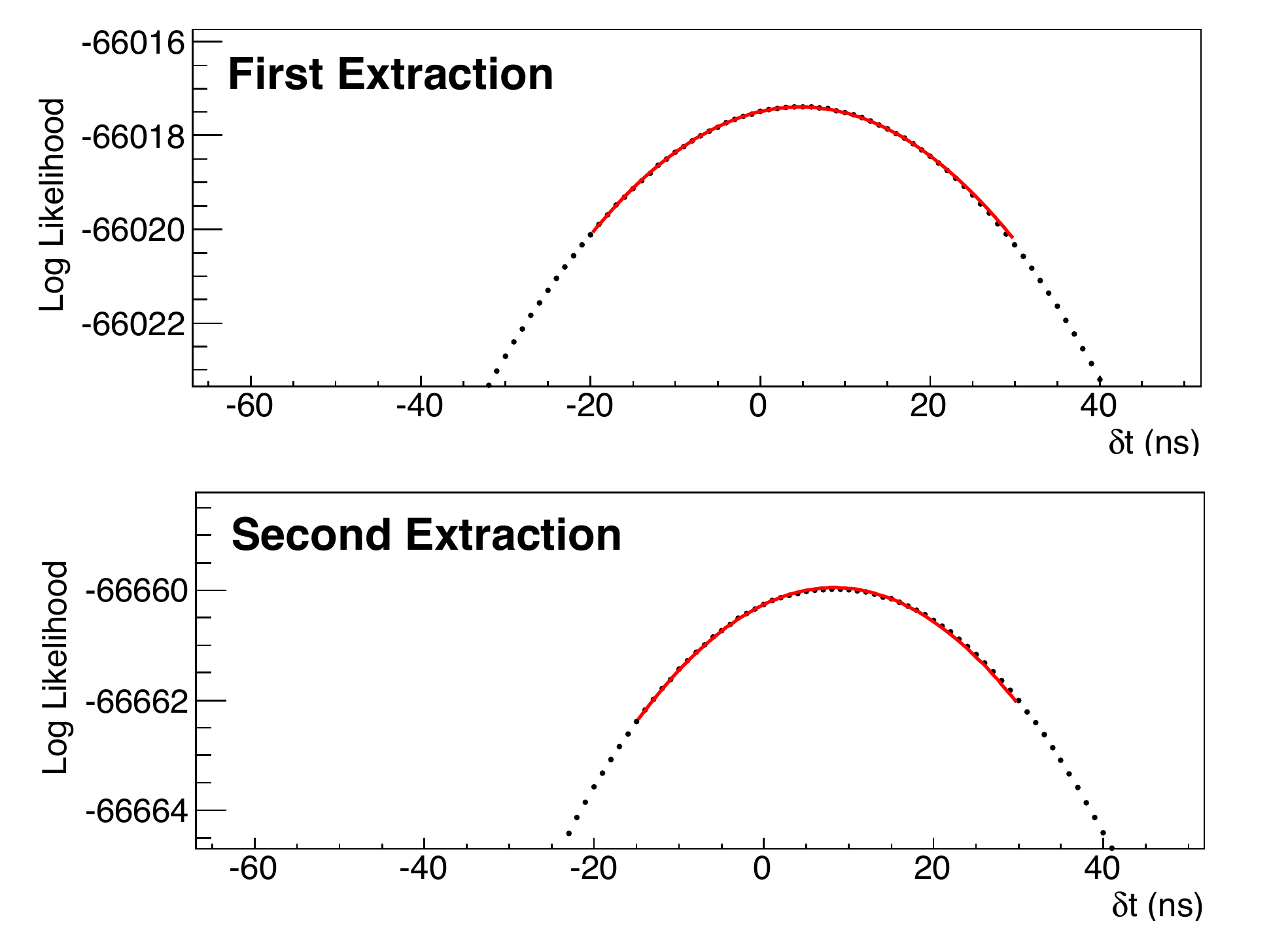}
 \caption{Log-likelihood distributions for both extractions as a function of ${\delta}$t, shown close to the maximum and fitted with a parabolic shape for the determination of the central value and of its uncertainty.}
 \label{fig10}
\end{figure}

\begin{figure}
\begin{center}
\resizebox{!}{5.2cm}{\includegraphics{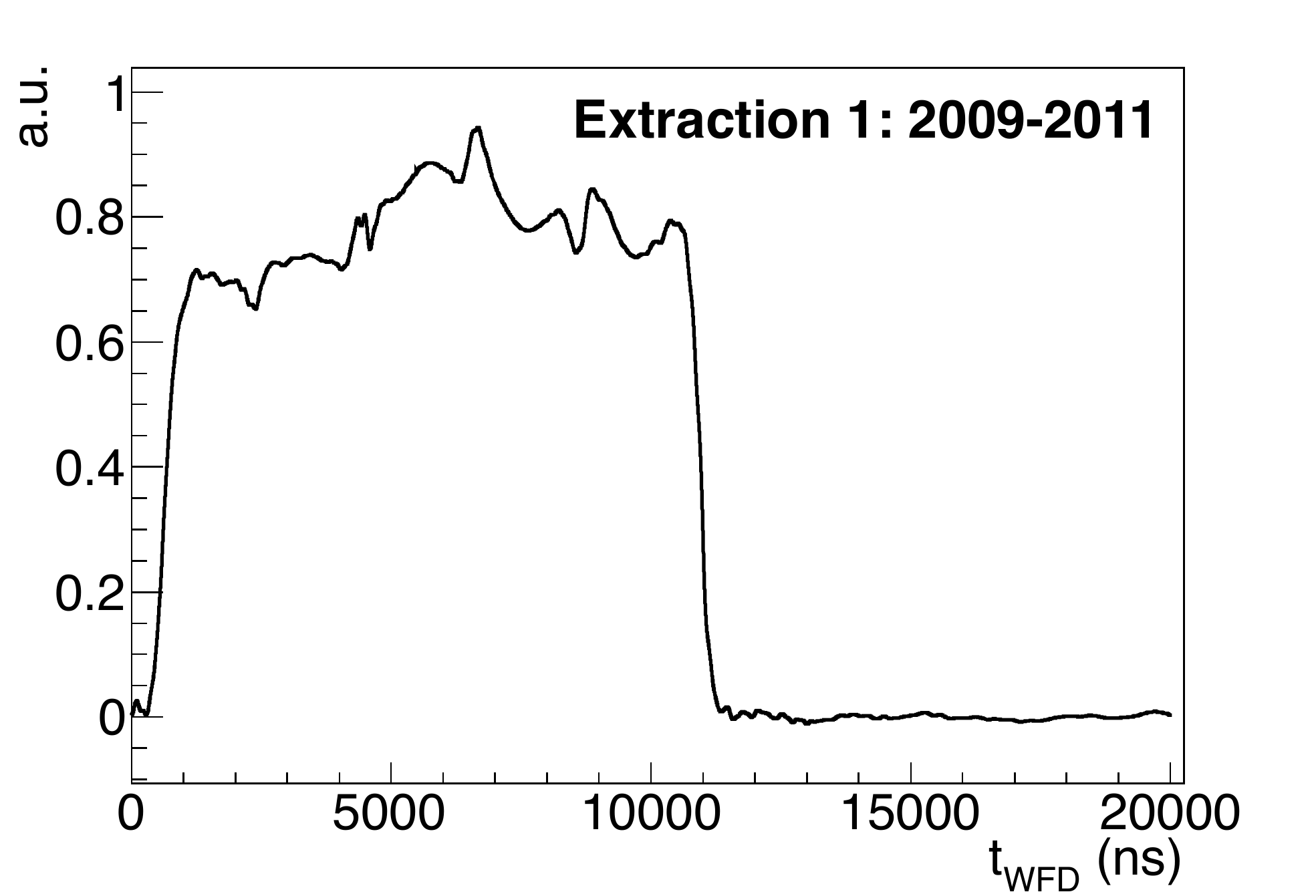}}
\resizebox{!}{5.2cm}{\includegraphics{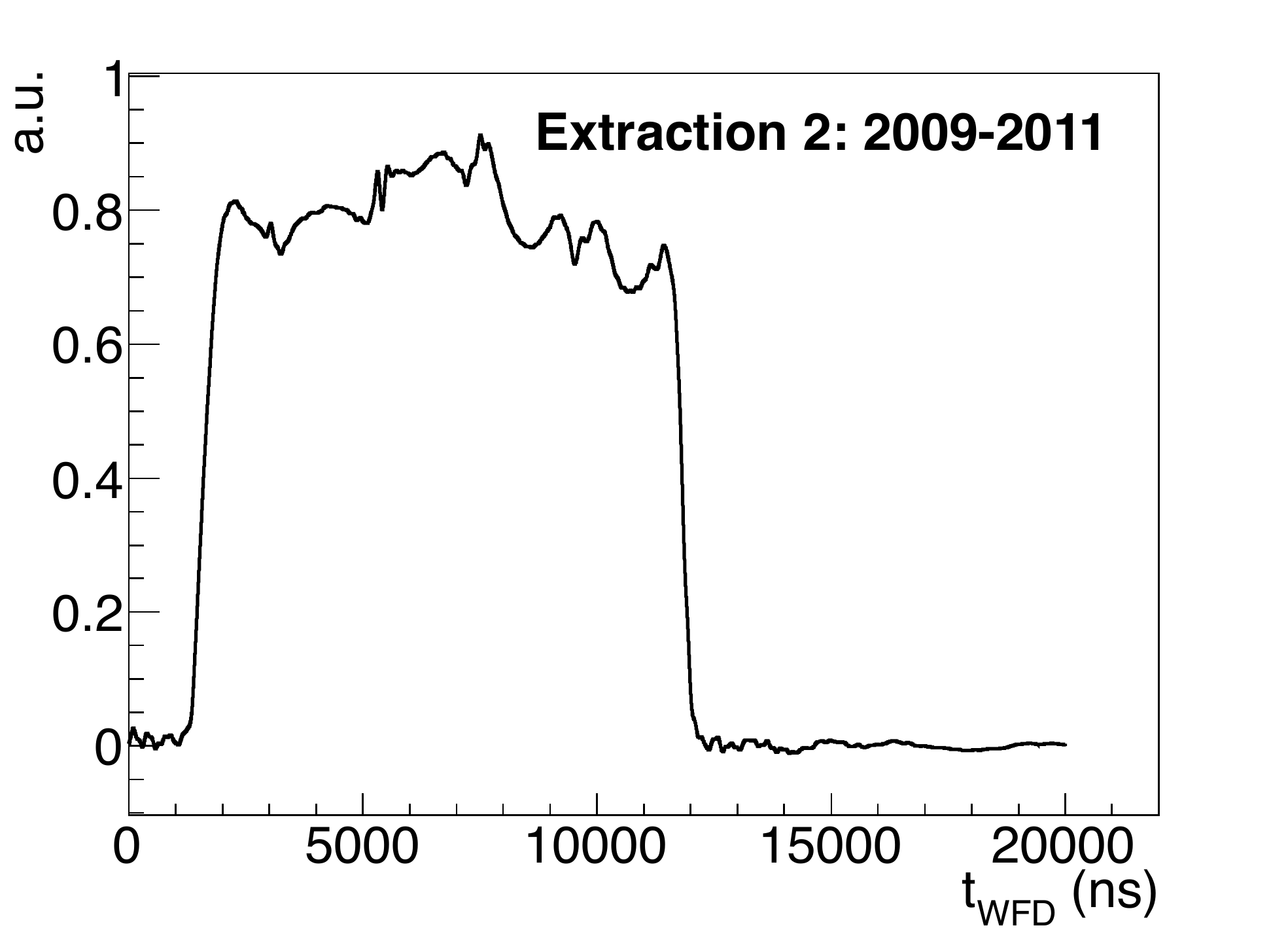}}
\end{center}
\caption{Summed proton waveforms of the OPERA events corresponding to the two SPS extractions for the 2009, 2010 and 2011 data samples.}
 \label{fig11}
\end{figure}

Near the maximum the likelihood function can be approximated by a Gaussian whose variance is a measure of the statistical uncertainty on $\delta$t. The data used for the maximum likelihood calculation are unbinned and the dependence on $\delta$t is computed by making a scan in steps of 1 ns. A parabolic fit is performed on the log-likelihood function for the evaluation of the maximum and of the statistical uncertainty (Fig.~\ref{fig10}). As seen in Fig.~\ref{fig11}, the PDF representing the time-structure of the proton extraction is not flat but exhibits a series of peaks and valleys, reflecting the features and the inefficiencies of the proton extraction from the PS to the SPS via the Continuous Transfer mechanism \cite{ref41}. Such structures may well change with time. The way the PDF are built automatically accounts for the beam conditions corresponding to the neutrino interactions detected by OPERA.

\begin{figure}
  \centering
  \includegraphics[height=.30\textheight]{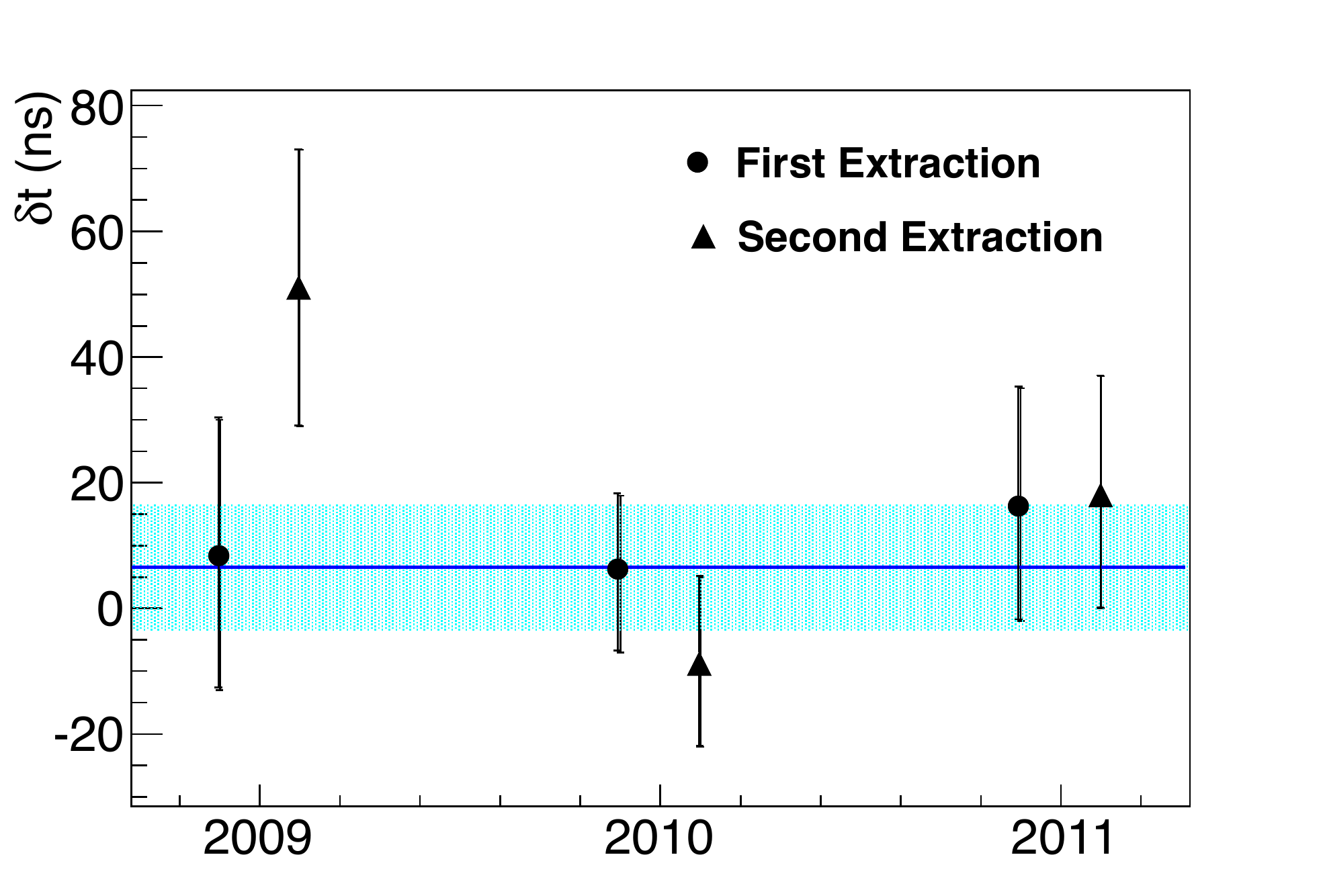}
  \caption{Results of the maximum likelihood analyses for ${\delta}$t corresponding to the two SPS extractions for the 2009, 2010 and 2011 data samples.}
 \label{fig12}
\end{figure}

The  results for $\delta t$ from the two proton extractions obtained for the years 2009, 2010 and 2011 are compared in Fig.~\ref{fig12}. They are compatible with each other. Data were also grouped in arbitrary subsamples to look for possible systematic dependencies. For example, by computing $\delta t$ separately for events taken during day (from 8 AM to 8 PM) and night hours, the absolute difference between the two calculations is ($16.4 \pm 15.8$)~ns providing no indication for a systematic effect.
In addition, with the presently available statistics we do not have indications of variations in the daily-24 hour observations. A similar result was obtained for a summer vs (spring plus autumn) dependence, possibly induced by thermal effects in the setup, which yielded ($15.6 \pm 15.0$)~ns. An analysis was also conducted by grouping events in two bins, corresponding to low- and high-intensity extractions (below and above $1.97\times 10^{13}$ protons on target, respectively). The absolute difference between the two bins is ($6.8 \pm 16.6$)~ns.

\ \ The maximum likelihood procedure was checked with a Monte Carlo simulation. Starting from the experimental PDF, an ensemble of 100 data sets of OPERA neutrino interactions was simulated. Data were shifted in time by a constant  value, hence faking a time of flight deviation. Each sample underwent the same maximum likelihood procedure as applied to real data. The analysis yielded a result accounting for the statistical fluctuations of the sample that are reflected in the different central values and their uncertainties. The average of the central values from this ensemble of simulated OPERA experiments reproduces well the time shift applied to the simulation (at the 0.3~ns level). The average statistical error extracted from the likelihood analysis also reproduces within 1~ns the RMS distribution of the mean values with respect to the true values.

\begin{figure}
\begin{center}
\resizebox{!}{5.1cm}{\includegraphics{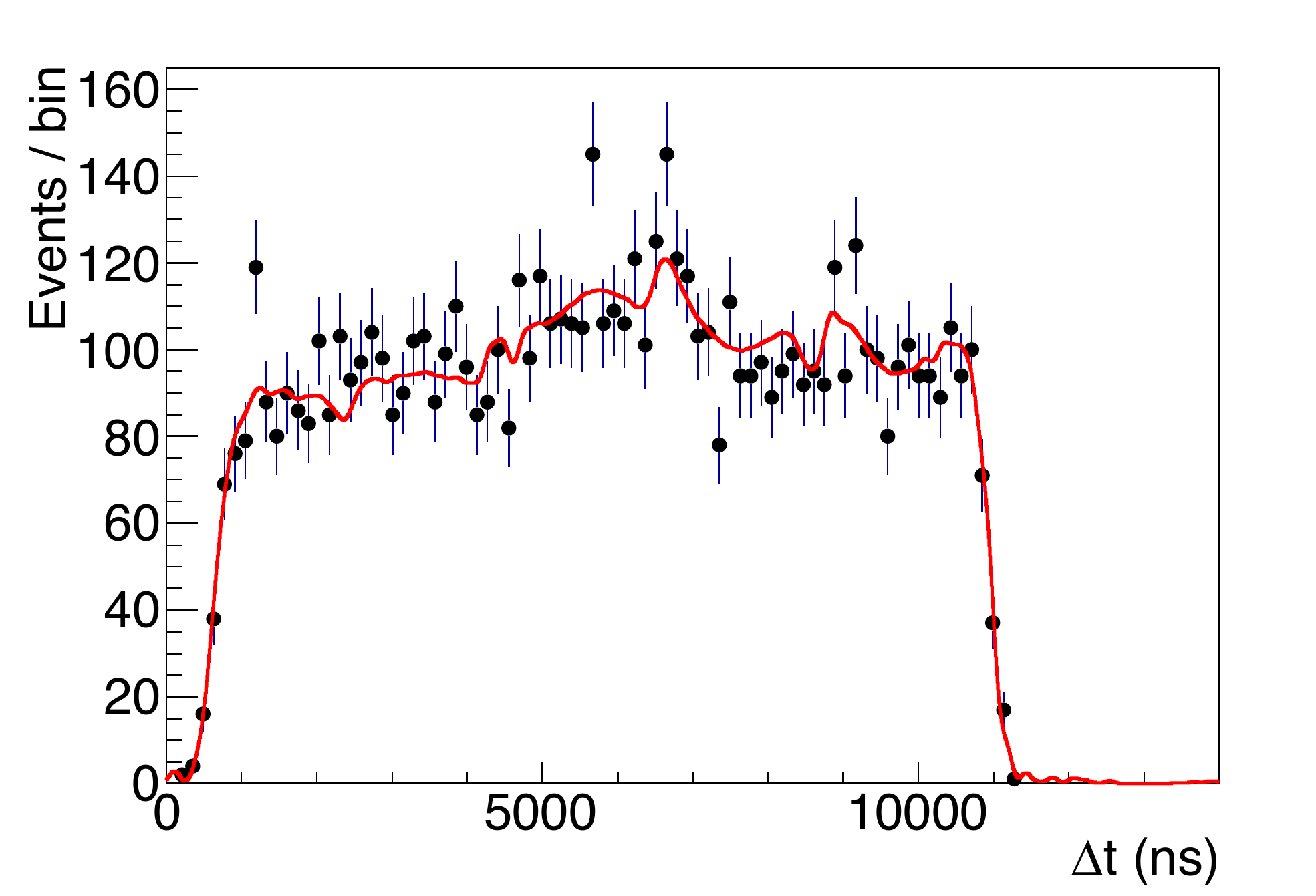}}
\resizebox{!}{5.1cm}{\includegraphics{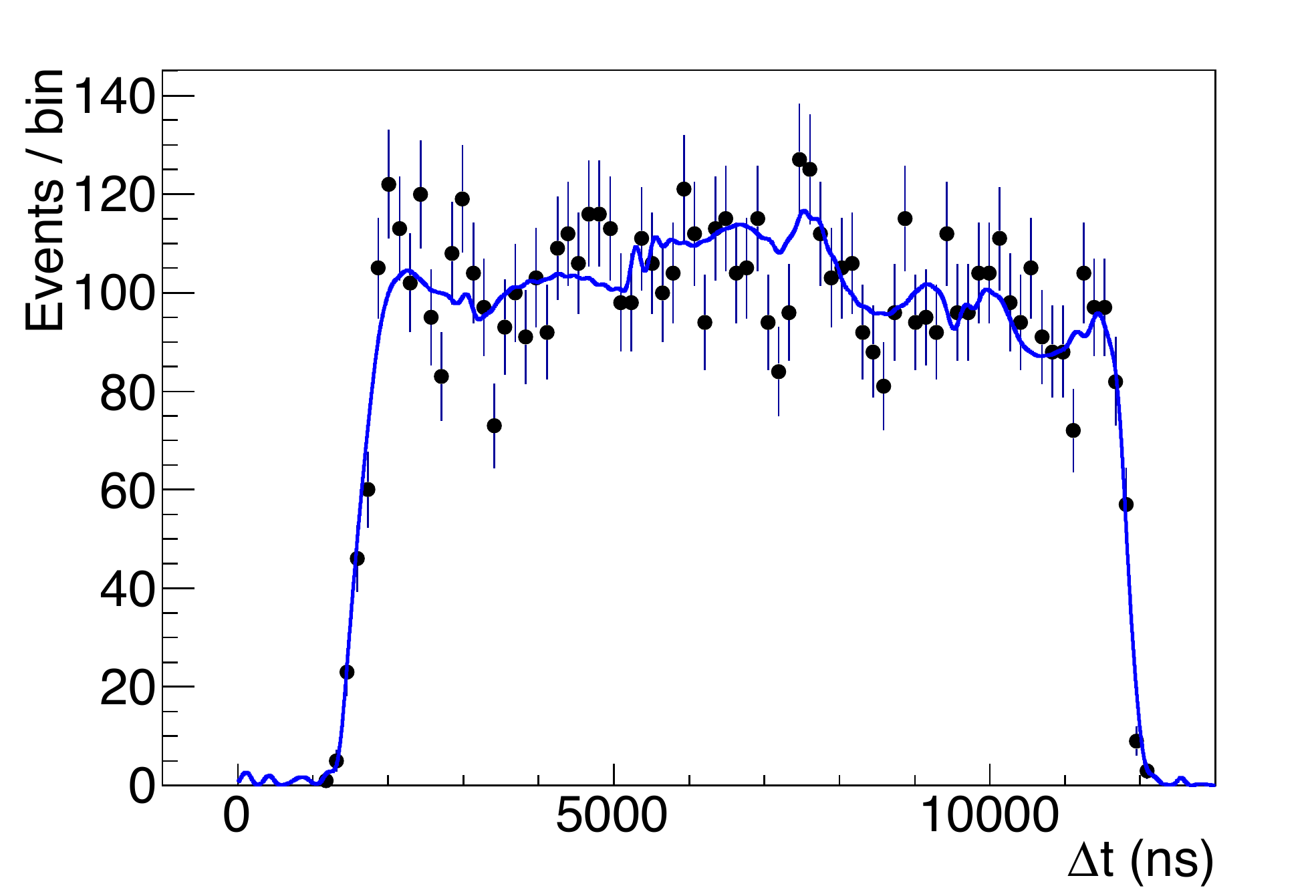}}
\end{center}
 \caption{Comparison of the measured neutrino interaction time distributions (data points) and the proton PDF (red and  blue line) for the two SPS extractions resulting from the maximum likelihood analysis.}
 \label{fig13}
\end{figure}

\begin{figure}
\vspace{0.3cm}
  \centering
  \includegraphics[height=.223\textheight]{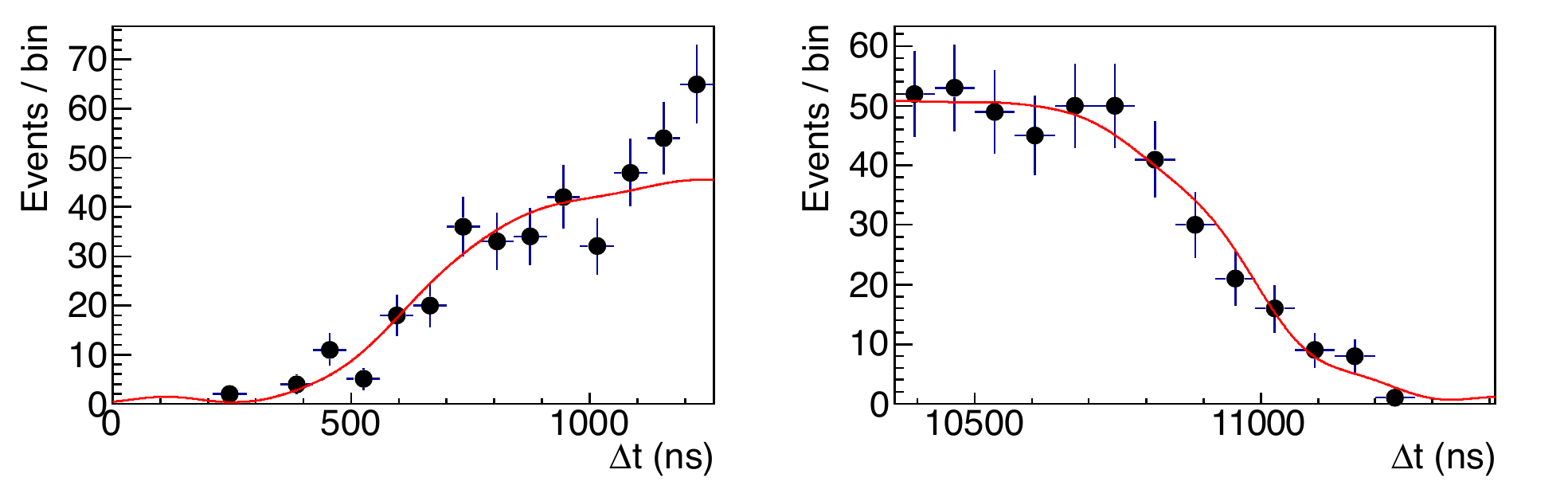}
  \caption{Blow-up of the leading (left plot) and trailing edge (right plot) of the measured neutrino interaction time distributions (data points) and the proton PDF (red line) for the first SPS extraction after correcting for $\delta t$=6.5 ns. Within errors the second extraction is exactly equal to the first one.}
 \label{fig14}

\end{figure}

The result of the  data analysis shows an arrival time of the neutrinos with respect to the one computed by assuming the speed of light:

{\centering
$$\delta t = TOF_c-TOF_\nu = (6.5 \pm 7.4~(stat.))~ns.$$
\par}

As a check, the same analysis was repeated considering only internal events. The result is  $\delta t = (14.2 \pm 11.9~(stat.))$~ns. The agreement between the proton PDF and the neutrino time distribution obtained after shifting by $\delta t$ is illustrated in Fig.~\ref{fig13}. Fig.~\ref{fig14} shows a blow-up of the leading and trailing edges of the distributions in Fig.~\ref{fig13}. In order to perform this comparison, data were binned and shifted by the value of $\delta t$ obtained from the maximum likelihood analysis. The plots in Figs.~\ref{fig13} and \ref{fig14} only give a visual representation of the agreement between the two binned distributions, which are not used for the extraction of the value of $\delta t$.

\begin{figure}
\centering
 \includegraphics[height=.33\textheight]{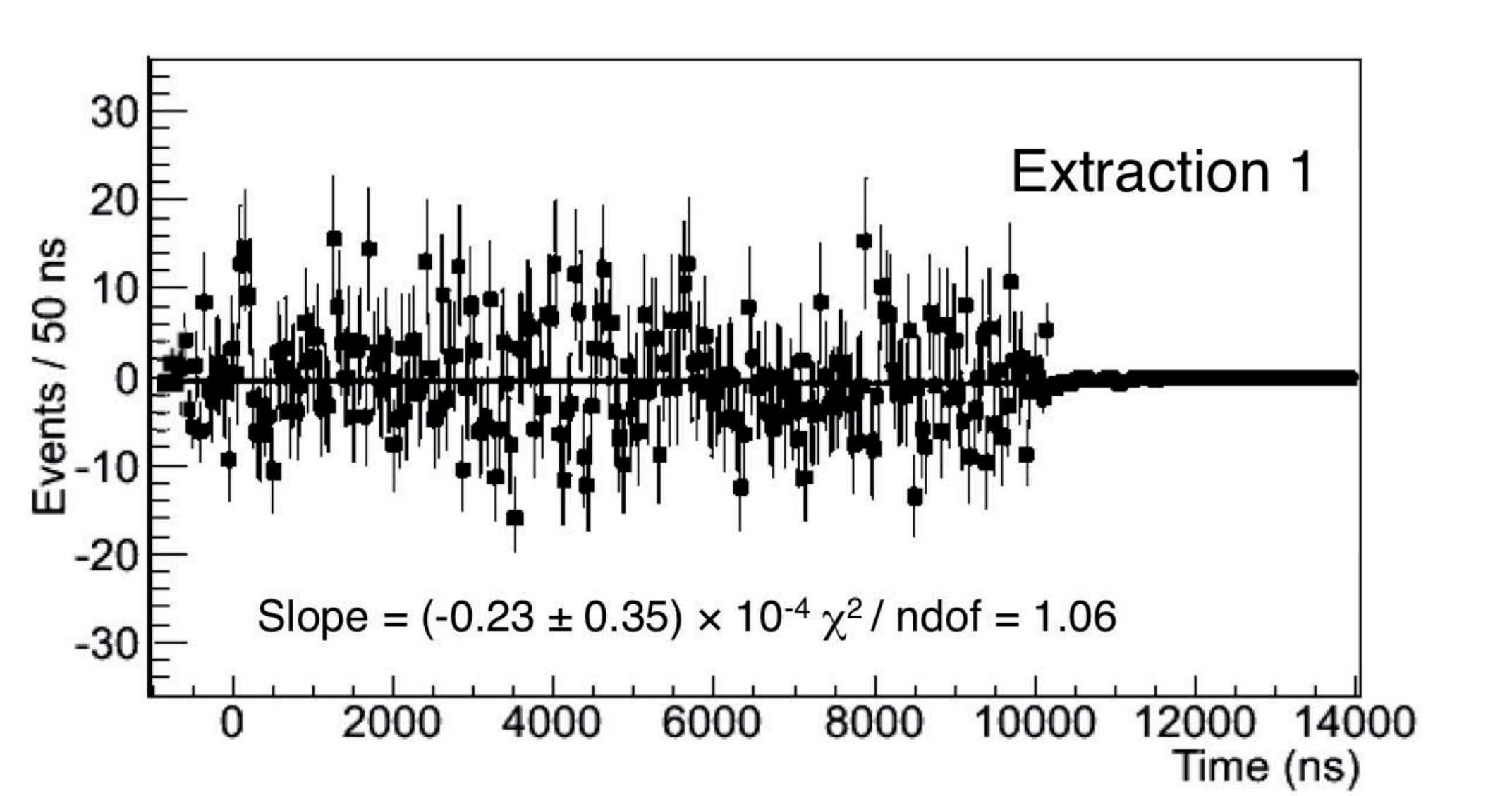}
 \caption{Residuals of the data points with respect to the PDF, after correcting for $\delta t$, as a function of the time since the start of the extraction. A linear fit superimposed to the experimental points gives results compatible with a flat distribution.  The slope value is given in number of events/ns.  Within errors the second extraction is exactly equal to the first one.}
 \label{fig15}
\end{figure}

The $\chi^2$/ndf for the full distribution is 1.1 for the first extraction and 1.0 for the second one.

Fig.~\ref{fig15} shows the residuals of the data points with respect to the PDF as a function of the time since the start of the extraction. No evidence is found for a time dependence.

\ \ Several additional statistical checks, such as the determination of $\delta t$ by a $\chi^2$ fit of the two distributions, or of parts of them separately (central region, leading and trailing edges) were performed to compare the proton PDF with the neutrino event distribution. These tests provided results comparable to those of the maximum likelihood analysis. None of them revealed any systematic effect within the present statistical accuracy, confirming the assumption that the neutrino event time distributions and the PDF are statistically equivalent.

\ \ An alternative method to extract the value of~$\delta t$ consists in building the likelihood~function by associating each neutrino interaction to its waveform instead of using the global PDF. This method can in principle lead to smaller statistical errors given the direct comparison of each event with its related waveform. However, particular care must be taken in filtering the electronic noise, white and coherent, that affects individual waveforms, while it cancels out in the global PDF. The two extractions can be~treated simultaneously in the same likelihood function defined as :

\begin{equation}
L (\delta t) = \prod_j W_j ( t_j + \delta t ) 
\end{equation}

Again, this procedure and the extraction of the statistical error were checked with a simulation. This method leads to a value of  $\delta$t = (3.5 $\pm$ 5.6~(stat.)) ns.

A systematic error of 4.4~ns is assigned to the different filtering methods and treatments of the waveform baseline. This error and the errors listed in Table~\ref{tab2}, add up to the total systematic uncertainty of this result given in Section~\ref{sec8}.

Finally, further investigations were conducted to search for possible systematic effects in the neutrino production mechanism by the SPS proton beam as measured by the BCT. The results are summarised below \cite{ref42}:

\begin{enumerate}
\item {
The neutrino production target contains 13 graphite rods, each 10 cm long. The first 8 are interspaced by 9 cm, the last 5 by 2 mm. The diametre of the first two rods is 5 mm; the other rods have a diametre of 4 mm. The proton transfer from the BCT to the target is practically lossless. The aiming accuracy to the target centre is within 50 (90) ${\mu}$m RMS on the horizontal (vertical) plane. The transport of the protons to the target does not introduce any acceptance effect on the neutrino yield and on the beam profile at LNGS, which is determined by the meson decay kinematics only.}

\item {
Density variations of the target during the time of extraction are negligible. The largest temperature increase in the graphite target corresponding to the point of maximum energy deposition (occurring around the second and the third rod) was estimated with a detailed simulation based on FLUKA and on a finite-element thermo-mechanical model of the target. This temperature increase corresponds on average to 297 K, yielding a density reduction due to transverse dilatation of 0.3\%, within the 10.5 ${\mu}$s duration of the extraction. The total target thickness is 3.3 interaction lengths. The local density variation at the point of maximum energy deposition translates into a small displacement of this point with a negligible effect on the number of interacting protons, neutrino yield and timing distribution.}

\item {
The current pulses of the horn and reflector magnets are 6.8 and 10 ms long, respectively; this is well above the proton extraction duration of 10.5 ${\mu}$s. The timing of the pulses is centred on the proton extractions and is continuously monitored. This timing is not critical with respect to the focalisation efficiency: tests were performed by artificially shifting the pulses of the magnets by as much as 100 ${\mu}$s. These extreme conditions produced a decrease of the muon flux associated to the neutrino beam by less than 1\%, confirming the focalisation stability with respect to the pulse timing under normal operating conditions.}

\end{enumerate}

\section{Results}
\label{sec8}

The  delay values used to derive the final $\delta t$ are summarised in Table~\ref{tab1}. 
One then obtains  $\delta t = TOF_c -TOF_\nu = (6.5 \pm 7.4~(stat.))$~ns. This result is also affected by an overall systematic uncertainty of (-8.0, +8.3) ns coming from the combination of the different terms already discussed and summarised in Table~\ref{tab2}. The total systematic uncertainty was computed numerically by taking into account the individual contributions and their corresponding probability distributions. The dominant uncertainty is due to the calibration of the BCT time response.
The error in the CNGS-OPERA GPS synchronisation has been computed by adding in quadrature the uncertainties on the calibration performed by PTB and the internal errors of the two high-accuracy GPS systems. The final systematic uncertainty is asymmetric. For external events, the position of the neutrino interaction in the rock is unknown and, in particular, its transverse position with respect to the detector. The distribution of the uncertainty on this position is flat. This systematically leads to an apparent increase of the neutrino time of flight $TOF_\nu$ and thus to a systematic decrease of $\delta t$.

 The final result of the measurement is then:

$$\delta t = TOF_c - TOF_\nu = (6.5 \pm 7.4\,~(stat.)~^{+8.3}_{-8.0}~(sys.))~ns$$

The relative difference of the muon neutrino velocity with respect to the speed of light is:

 $$(v-c)/c = \delta t /(TOF'_c - \delta t) = (2.7 \pm 3.1~(stat.)~^{+3.4}_{-3.3}~(sys.)) \times 10^{-6}$$

In performing this last calculation a baseline of 730.085~km was used, and $TOF'_c$ corresponds to this effective neutrino baseline starting from the average meson decay point in the CNGS-CERN tunnel as determined by simulations. Actually, the $\delta t$ value is measured over the distance from the BCT to the OPERA reference frame, and it is only determined by neutrinos and not by charged pions and kaons, which introduce negligible delays.

\begin{table}
  \centering
\caption{Contribution to the overall systematic uncertainty on the measurement of $\delta t$.}
\begin{tabular}{lrc}
\hline
 \bf{Error source} & \bf{ns} & \bf{Error distribution} \\
\hline  
Baseline (20~cm)   & 0.67   & Gaussian \\
Meson decay point             & 0.2     & Exponential (1~side) \\
Interaction point  of external neutrino events    & 2.0    & Flat (1~side) \\
&\\
CTRI signal propagation through GMT chain   $\Delta t_{UTC}$             & 2.0    & Gaussian \\
Kicker magnet signal to WFD $\Delta t_{trigger}$   & 1.0 & Gaussian \\&\\
BCT calibration  $\Delta t_{BCT}$  & 5.0 & Gaussian \\

CNGS-OPERA GPS synchronisation   & 1.7 & Gaussian \\&\\
 
LNGS 8.3 km fibre to OPERA Master Clock              & 3.7 & Gaussian \\
Master Clock to FPGA $\Delta t_{clock}$   & 1.0 & Gaussian \\
TT PMT signal to FPGA   & 2.3 & Gaussian \\
TT timing simulation of $\nu$ interactions  & 3.0 & Gaussian \\
FPGA latency   & 1.0 & Gaussian \\

 & & \\
RPC signal formation & 5.0 & Gaussian \\
RPC signal propagation & 1.0 & Gaussian   \\
RPC FEB delay  & 1.0 & Gaussian  \\
RPC DAQ clock trasmission & 2.0 & Gaussian\\
RPC FPGA calibration & 1.0 & Gaussian\\
RPC plane disuniformity & 3.9 & Gaussian\\ 
\hline
\bf{ Total systematic uncertainty for TT based analysis}   & $-8.0, +8.3$&  \\
\bf{ Total systematic uncertainty for RPC based analysis}   & $-9.6, +9.9$&  \\
\hline
\end{tabular}
\label{tab2}
\end{table}


\ \ The alternative analysis in which the likelihood~function is built by associating each neutrino interaction to its waveform instead of using the global PDF leads to a compatible value  of $\delta t= (3.5 \pm 5.6~(stat.)^{+9.4}_{-9.1} (sys.))$ ns. The systematic uncertainty includes the additional contribution of 4.4 ns resulting from more complex noise filtering and baseline treatment of the waveforms. 

The dependence of $\delta t$ on the neutrino energy was also investigated, yielding a null result.

\begin{figure}
  \centering
  \includegraphics[height=.35\textheight]{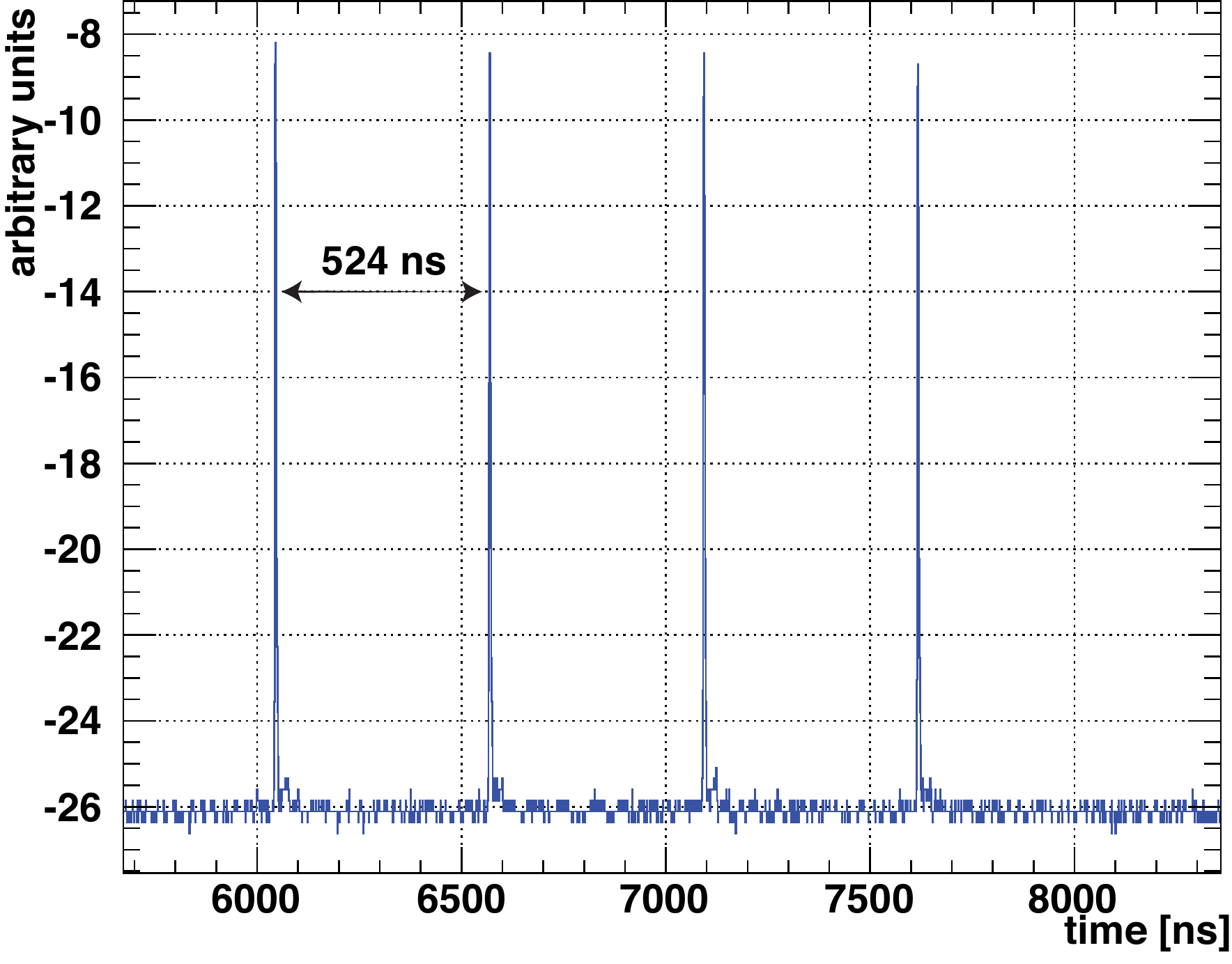}
  \caption{Timing structure of the four-bunch proton extraction of the dedicated CNGS bunched neutrino beam as read out by the BCT detector.}
 \label{fig16}
\end{figure}

\begin{figure}
  \centering
  \includegraphics[height=.33\textheight]{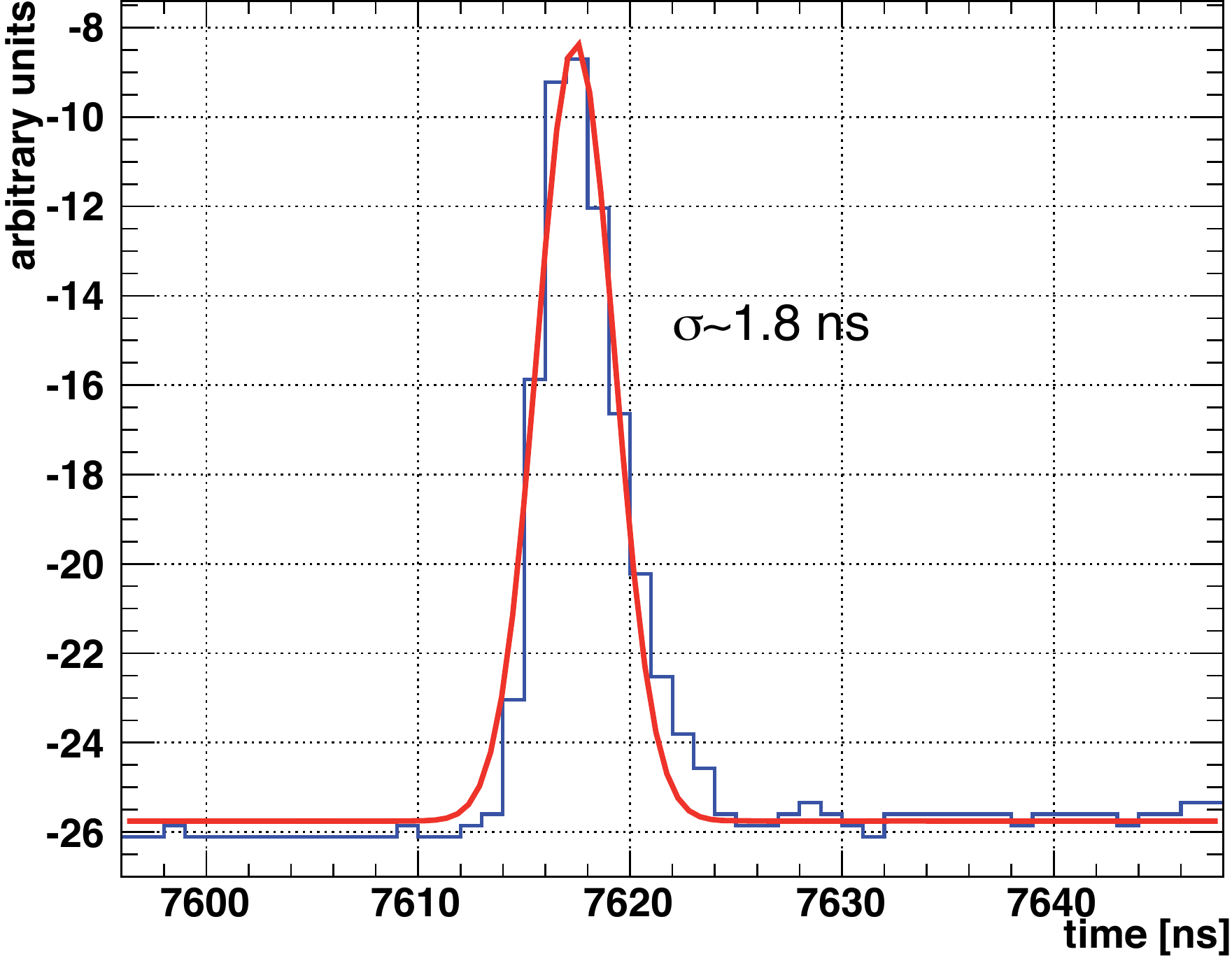}
  \caption{Timing structure for one individual proton bunch (folded with the BCT time response).}
 \label{fig17}
\end{figure}

\section{Test with a short-bunch wide-spacing beam}
\label{sec9}

\ \ In order to exclude possible systematic effects related to the use of the proton waveforms as PDF for the distributions of the neutrino arrival times within the two extractions and to their statistical treatment, a test was performed with a dedicated CNGS beam generated by  a proton beam set up on purpose for the neutrino velocity measurement. The modified SPS super-cycle consisted of a single extraction including four bunches about 3 ns long (FWHM) separated by 524 ns, yielding a total of $1.1\times 10^{12}$ protons per cycle. One typical proton extraction read out by the BCT is shown in Fig.~\ref{fig16}, while Fig.~\ref{fig17} shows an expanded view of an individual bunch waveform. This beam is similar to the one used for the BCT calibration discussed in Section~\ref{sec6} and it allowed performing time of flight measurements at the single event level.

\ \ Running with the CNGS bunched beam lasted from October 22 to November 6, 2011 for a total integrated intensity of $4\times 10^{16}$ protons on target. A total of 35 beam-related neutrino events were collected by OPERA. The events were then selected and reconstructed in the same way as those used for the main analysis. After selection, 6 internal and 14 external events were retained. Within the small statistics the events are evenly distributed in the four bunches of the extraction.

\ \ Given the short bunch length and the relatively long inter-bunch distance one could unambiguously associate each neutrino event to its corresponding proton bunch. The price to pay for achieving such a high definition of the neutrino emission time is the very low beam intensity, on the average about 60 times lower than for normal CNGS operation.

\subsection{Results with the Target Tracker data}
\label{sec:tt}

\begin{figure}
  \centering
  \includegraphics[height=.35\textheight]{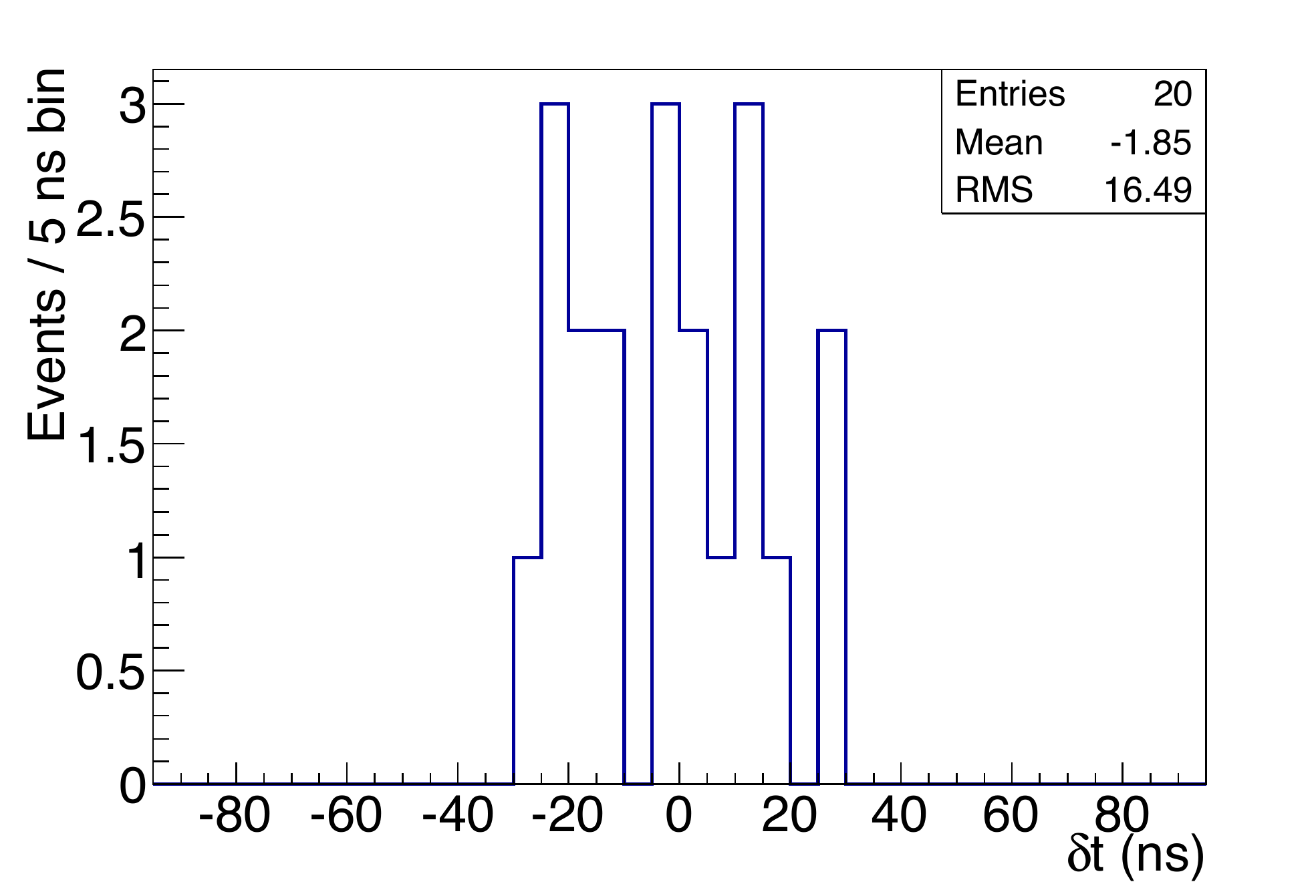}
  \caption{Distribution of the $\delta t$ values obtained from the 20 TT events taken with the bunched neutrino beam. The mean value is  ($-1.9 \pm 3.7$)~ns.}
 \label{fig18}
\end{figure}

 Fig.~\ref{fig18} shows the distribution of the values of $\delta t$ obtained for the events detected by the TT during the bunched beam test. The RMS is 16.5 ns and the average is ($-1.9 \pm 3.7$) ns in agreement with the value of  ($6.5 \pm 7.4$) ns obtained with the main analysis; it is also in agreement with the ICARUS result reported in \cite{arXiv-Icarus}. At first order, systematic uncertainties related to the bunched beam operation are equal or smaller than those affecting the result with the nominal CNGS beam. The main contributions to the dispersion are given by the TT response of 7.3~ns RMS, the DAQ time granularity of 10~ns full width, and the jitter of $\pm~25$~ns related to the tagging of the external GPS signal by the OPERA Master Clock. The latter dominant term results in a RMS of 14.4~ns \mbox{(50 ns/${\surd}$12)}. This dispersion is only relevant for the bunched beam measurement; nevertheless the statistical accuracy on the average ${\delta}$t is already as small as 3.7~ns with only 20 events.

\ \ This result largely excludes possible biases affecting the statistical analysis based on the proton PDF. It also indicates the absence of significant biases due to the cumulative response of the beam line to long proton pulses (target aiming accuracy, horns' timing, target temperature increase), as well as pulse duration effects in the BCT response. Moreover, since waveform filtering does not apply, the above considerations concern as well the procedure adopted for removing the noise.\\

\subsection{Results with the RPC data}

\label{sec:rpc}

Each of the two OPERA magnets is instrumented with 22 planes of Resistive Plate Chambers (RPC)  \cite{ref1}. In each plane, copper readout strips measuring the electric signal generated by the crossing of a charged particle through the gas gap provide horizontal and vertical coordinates in the detector transverse plane. In addition, the DAQ records for each plane the time stamp of the earliest signal reaching the readout electronics. An independent time calibration of the RPC was performed in order to translate the time measured locally in UTC time.
\par With the magnetic spectrometers the momentum and the charge sign of the high energetic muons (mainly from $\nu_\mu$ CC interactions) leaving the target are measured. Hits not attached to a muon track are ignored as they can be due to noise or particles out of time with respect to the neutrino interaction. Every RPC plane provides an independent time measurement along each muon track at a known position along the detector. 
This sequence of independent time measurements, corrected for all delays, can then be time translated to the origin of the OPERA reference frame and averaged.
 As the number of measurements ranges between about  20 to about 40 for a track crossing both Super Modules the 10 ns DAQ time quantisation affecting single measurements is washed out. For tracks crossing the full detector, two independent measurements can be extracted, one for each Super Module separately. The distribution of the difference between the two mean times was obtained for muon tracks collected during the full 2011 run. Its mean is 0.4 ns with an RMS of 5.5 ns.
\par The outlined procedure was applied to all CC events of the 2011 bunched beam run for which a neutrino interaction time could be evaluated using the RPC data. The $\delta t$ distribution is shown in Fig.~\ref{fig:deltat-rpc}. All the events fall within the $\pm$~25~ns time interval expected from the jitter on the GPS signal tagging. The value of $\delta t$ obtained from the RPC data is:

$$\delta t = (-0.8~\pm~3.5~(stat.)^{+9.9}_{-9.6} (sys.))~ns.$$

The systematic error takes into account those uncertainties listed in Table 2 that affect the RPC data.

\bfi[htbp]
\bc
\includegraphics[width=12.cm]{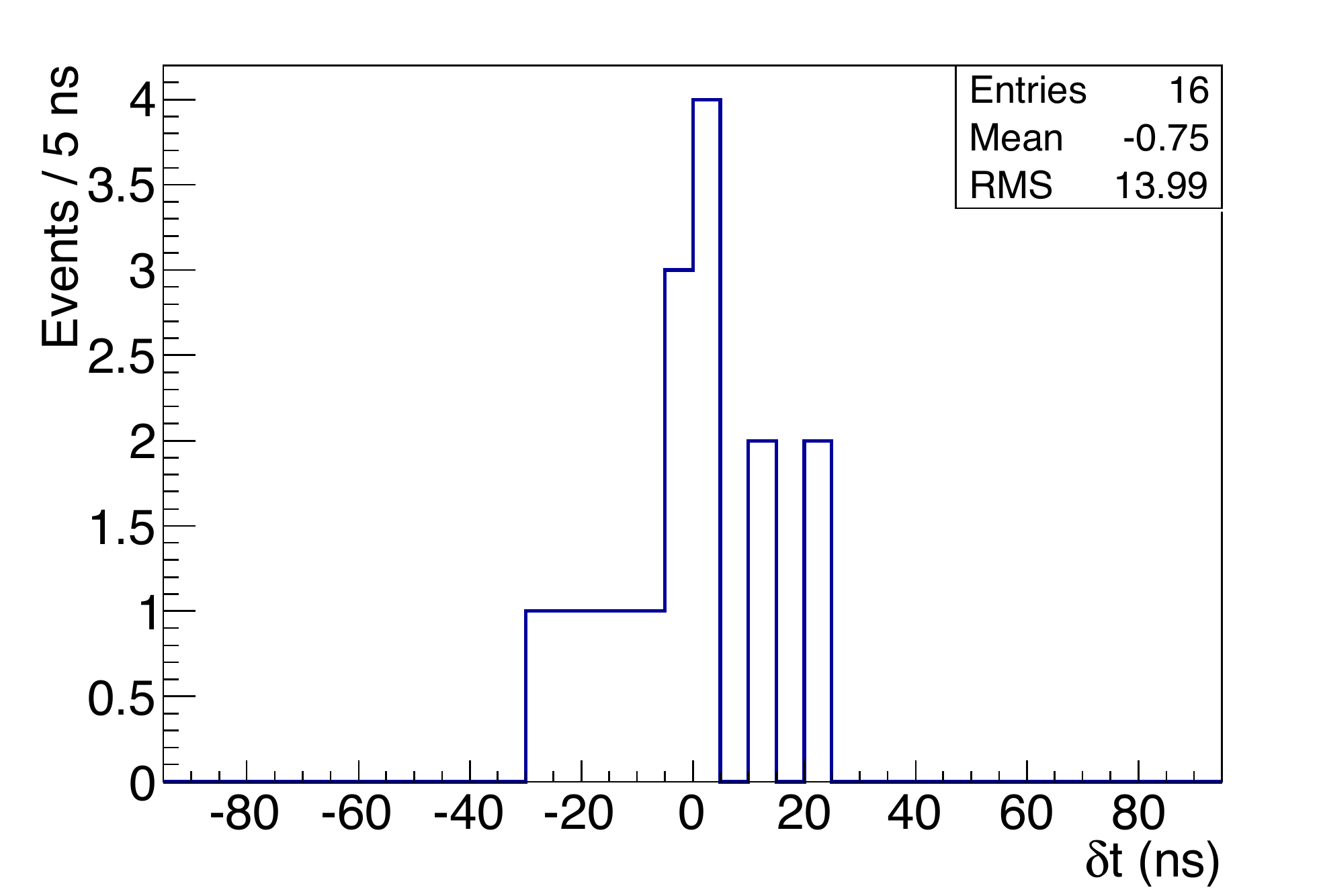}
\caption{$\delta t$ distribution of 16 events obtained from the analysis of the RPC data.}
\label{fig:deltat-rpc}
\ec
\efi

\section{Conclusions}
\label{sec10}

The OPERA detector at LNGS, designed for the study of neutrino oscillations in appearance mode, provided a precision measurement of the neutrino velocity over the 730 km baseline of the CNGS neutrino beam sent from CERN to LNGS through the Earth's crust. A time of flight measurement with small systematic uncertainties was made possible by a series of accurate metrology techniques pionereed by the OPERA Collaboration. The data analysis took also advantage of a large sample of 15223 neutrino interaction events detected by OPERA.

The analysis of internal neutral current and charged current events, and external $\nu_\mu$ CC interactions from the 2009, 2010 and 2011 CNGS data was carried out to measure the neutrino velocity, $v$. The sensitivity of the measurement of ($v-c)/c$ is about one order of magnitude better than former accelerator neutrino experiments.

The results of the study using CNGS muon neutrinos with an average energy of 17 GeV indicate a neutrino arrival time  compatible within errors to the one computed by assuming the speed of light in vacuum:

 $$\delta t = (6.5 \pm 7.4~(stat.)^{+8.3}_{-8.0} (sys.))~ns.$$

The corresponding relative difference of the muon neutrino velocity and the speed of light is:

$$(v-c)/c = (2.7 \pm 3.1~(stat.)^{+3.4}_{-3.3} (sys.)) \times 10^{-6}.$$


\ \ An alternative analysis in which the likelihood~function is built by associating each neutrino interaction to its waveform instead of using the global PDF leads to a compatible value of $\delta t = (3.5 \pm 5.6~(stat.)^{+9.4}_{-9.1} (sys.))$~ns affected by an additional contribution to the systematic error.

The dependence of $\delta t$ on the neutrino energy was also investigated yielding a null effect.

To exclude possible systematic effects related to the use of the proton waveforms as PDF for the distributions of the neutrino arrival times within the two extractions and to their statistical treatment, a two-week long beam test was performed at the end of 2011.
A dedicated CNGS beam was generated by  an SPS proton beam set up for the purpose of the neutrino velocity measurement. The modified beam consisted of a single extraction including four bunches about 3 ns long (FWHM) separated by 524 ns. With an integrated beam intensity of $4\times 10^{16}$ protons on target a total of 20  TT and 16 RPC events were retained, leading to a value of $\delta t$ measured from the average of the TT distribution of ($-1.9\pm 3.7$)~ns  and ($-0.8\pm 3.5$)~ns from the RPC, in agreement with the value of ($6.5 \pm 7.4$)~ns obtained with the  main analysis. At first order, systematic uncertainties related to the bunched beam operation are equal or smaller than those affecting the result obtained with the  standard CNGS beam.\par
After several months of additional studies, with the new results reported in this paper, the OPERA Collaboration has completed the scrutiny of the originally reported neutrino velocity anomaly by identifying its instrumental sources and coming to a coherent interpretation scheme.

\section{Acknowledgements}
\label{sec11}

We thank CERN for the successful operation of the accelerator complex and the CNGS facility, and for the prompt setting up of the bunched proton beam. We are indebted to INFN for the continuous support given to the experiment during the construction, installation and commissioning phases through its LNGS laboratory. Funding from our national agencies is warmly acknowledged: Fonds de la Recherche Scientifique - FNRS and Institut Interuniversitaire des Sciences Nucleaires for Belgium; MoSES for Croatia; CNRS and IN2P3 for France; BMBF for Germany; INFN for Italy; JSPS (Japan Society for the Promotion of Science), MEXT (Ministry of Education, Culture, Sports, Science and Technology), QFPU (Global COE program of Nagoya University, ''Quest for Fundamental Principles in the Universe'' supported by JSPS and MEXT) and Promotion and Mutual Aid Corporation for Private Schools of Japan for Japan; The Swiss National Science Foundation (SNF), the University of Bern and ETH Zurich for Switzerland; the Russian Foundation for Basic Research(grant 09-02-00300 a), the Programs of the Presidium of the Russian Academy of Sciences ``Neutrino Physics'' and ``Experimental and theoretical researches of fundamental interactions connected with work on the accelerator of CERN'', the Programs of support of leading schools (grant 3517.2010.2), and the Ministry of Education and Science~ of~ the Russian Federation for Russia (contract 12.741.12.0150); the Korea Research Foundation Grant (KRF-2008-313-C00201) for Korea; and TUBITAK The Scientific and Technological Research Council of Turkey, for Turkey. We are also indebted to INFN for providing fellowships and grants to non-Italian researchers. We thank the IN2P3 Computing Centre (CC-IN2P3) for providing computing resources for the analysis and hosting the central database for the OPERA experiment. We gratefully acknowledge the support of the LNGS Computing Centre and of the INFN-Bologna Electronic Lab. We are indebted to our technical collaborators for the excellent quality of their work over many years of design, prototyping and construction of the detector and of its facilities.

\end{document}